\documentclass[12p]{article}
\usepackage{latexsym}
\usepackage{color}
\usepackage{amsmath}
\usepackage{epsfig}
\usepackage{hyperref}
 \usepackage{subfigure}
 \usepackage{dcolumn}
   \usepackage{threeparttable}

\newcommand{\ortala}[1]{\begin{center}#1\end{center}}

\newcommand{\sandd}[1]{\left\langle #1\right\rangle}

\newcommand{\integ}[3]{{{\underset{#1 }{\overset{#2}{\displaystyle\int}}}#3}}

\newcommand{\summ}[3]{{{\underset{#1 }{\overset{#2}{\displaystyle\sum}}}#3}}

\newcommand{\re}[1]{(\ref{#1})}

\newcommand{\eq}[2]{\begin{equation}\label{#1}  #2\end{equation}}

\newcommand{\paran}[1]{\left(#1\right)}
\newcommand{\mutlak}[1]{\left|#1\right|}

\newcommand{\sch}[1]{Schrodinger}

\newcommand{\komb}[2]{\paran{\begin{array}{c} #1 \\ #2 \end{array}}}

\usepackage{amssymb}
\usepackage{latexsym}
\begin{document}

\ortala{\large\textbf{Effects of the randomly distributed magnetic field on the phase diagrams of the transverse Ising thin film}}

\ortala{\textbf{\"Umit Ak\i nc\i \footnote{\textbf{umit.akinci@deu.edu.tr}}}}

\ortala{\textit{Department of Physics, Dokuz Eyl\"ul University,
TR-35160 Izmir, Turkey}}

\section{Abstract}

The effect of the zero centered Gaussian random magnetic field distribution on the phase
diagrams and ground state magnetizations of the transverse Ising thin film  has been investigated.  As a formulation, the differential operator technique and decoupling approximation within the  effective field theory has been used. The variation of the phase diagrams with the Gaussian distribution width ($\sigma$) has been obtained and particular attention has been paid on the evolution of the special point coordinate with distribution parameter. In addition, the ground state longitudinal and transverse magnetization  behaviors have been investigated in detail.

\section{Introduction}\label{introduction}

Recently there has been growing interest both theoretically and
experimentally on the finite magnetic materials especially on semi-infinite systems and thin films, since the
magnetic properties of free surfaces are drastically different from the bulk material, because the free surface breaks the translational symmetry, i.e.
surface atoms are embedded in an environment of lower symmetry than that of the inner atoms \cite{ref1,ref2}. If the surface exchange coupling is greater than a critical value, the surface region can exhibit an ordered phase even if the bulk is paramagnetic and it has a transition temperature higher than the bulk one. This fact has been observed experimentally\cite{ref3,ref4,ref5}.

Also, the development of the molecular beam epitaxy technique and its application to the growth of thin metallic films has stimulated renewed interest in thin film magnetism. It was experimentally found that, the Curie temperature and the average magnetic moment per atom increases with the increasing thickness of the film \cite{ref6,ref7}. Thin films can be modeled by Ising model due to the fact that, many ultrathin films do indeed exhibit a strong uniaxial anisotropy\cite{ref8}. These systems have been widely studied in literature by means of several theoretical methods such as Monte Carlo (MC) simulations \cite{ref9,ref10,ref11,ref12,ref13,ref14},  mean field approximation (MFA) \cite{ref15,ref16} and effective field theory (EFT) \cite{ref17,ref18}.

But the study of ferroelectric films treated by transverse Ising model (TIM) are more common in the literature. Oxide thin films with perovskite-type structure (such as  Barium titanate $BaTiO_3$ and Strontium titanate  $SrTiO_3$) have been fabricated \cite{ref19,ref20,ref21} and novel physical properties found in these ferroelectric thin films which have many application areas in technology \cite{ref22}, such as optoelectronics \cite{ref23}, microelectronics \cite{ref24} and random access memory \cite{ref25}.
Theoretically, the TIM has often been applied for the investigation of these systems. TIM was originally introduced by Gennes et al. \cite{ref26} to describe the bulk phase transition of hydrogen-bonded ferroelectrics such as $KH_2PO_4$. Since then, this model has been applied to several physical systems like $DyVO_2$. After then, this model has been extended to study of several other systems, such as Jahn–Teller and ferromagnetic systems\cite{ref27} treated by means of a wide variety of techniques including MFA \cite{ref28,ref29,ref30,ref31,ref32,ref33,ref34} and EFT \cite{ref35,ref36,ref37,ref38,ref39}. Effect of the dilution on the phase diagrams and magnetic properties of the ferroelectric thin films (by means of the TIM) has been studied within the EFT formulation \cite{ref40,ref41,ref42,ref43,ref44}. Several authors have also taken into account the long range interactions \cite{ref45,ref46,ref47}. In order to simulate more realistic models, different exchange interactions have been defined and different transverse fields have been considered on the surface and bulk, which anticipates the mimics of the surface effects within the framework of EFT \cite{ref48,ref49,ref50,ref51,ref52,ref53,ref54,ref55}. Another improvement of the Ising thin film models is the consideration of amorphisation of the surface  due to environmental effects, and these realizations have also been solved with using EFT \cite{ref56,ref57,ref58,ref59,ref60,ref61}. Moreover, in order to mimic the effect of some imperfections on the film characteristics, straight line of magnetic adatoms \cite{ref62}, defect layers \cite{ref63} and seeding layer effects \cite{ref64} have been studied. Apart from these works, decorated transverse Ising thin films with EFT \cite{ref65,ref66,ref67,ref68} and MC \cite{ref69} have been studied. There are also higher spin Ising thin films e.g. spin-1 Ising thin films have been studied \cite{ref70,ref71}.

In the semi-infinite systems, depending on the ratio between surface exchange interaction and bulk exchange interaction, the system may order on the surface before it orders in the bulk which is called extraordinary transition. In the contrary of this, ordinary transition means that the surface critical temperature is the same as the bulk transition temperature. In the phase diagram, the intersection point between these two transitions is called special point. As the film gets thicker, it approaches the semi-infinite system; this unusual effect shows itself as an intersection point in the phase diagrams of the films that have different thickness plotted in critical temperature versus surface exchange interaction plane.

On the other hand, the Ising model in a quenched random field (RFIM) has been studied
over three decades. The model which is actually based on the local
fields acting on the lattice sites which are taken to be random
according to a given probability distribution was introduced for the
first time by Larkin \cite{ref72} for superconductors and later
generalized by Imry and Ma \cite{ref73}. Beside the similarities
between diluted antiferromagnets in a homogenous magnetic field and ferromagnetic systems in the presence of random
fields \cite{ref74,ref75}, a rich class of experimentally accessible disordered systems can be described by RFIM such as structural
phase transitions in random alloys, commensurate charge-
density-wave systems with  impurity  pinning, binary fluid
mixtures  in  random  porous  media,  and  the  melting  of
intercalates  in  layered compounds such as $TiS_2$\cite{ref76}. RFIM generally  models  the phase  transitions  and  interfaces  in  random  media \cite{ref77,ref78}, e.g prewetting transition on a disordered substrate can be mapped to  2D RFIM problem\cite{ref79}.

The aim of this work is the determine the Gaussian random longitudinal magnetic field distribution on the phase diagrams and magnetization behavior of the transverse Ising thin film. For this aim, the paper is organized as follows: In Sec. \ref{formulation} we
briefly present the model and  formulation. The results and
discussions are presented in Sec. \ref{results}, and finally Sec.
\ref{conclusion} contains our conclusions.

\section{Model and Formulation}\label{formulation}

Thin film can be modeled by a layered structure which consist of interacting $L$ parallel layers. Each layer is defined as a regular lattice with coordination number $z$.
The Hamiltonian of the thin film is given by
\eq{denk1}{\mathcal{H}=-\summ{<i,j>}{}{J_{ij}s_i^zs_j^z}-\Omega \summ{i}{}{s_i^x}-\summ{i}{}{H_is_i^z}}
where $s_i^z$ is the $z$ component of the spin-$\frac{1}{2}$ operator (which can take the values $s_i^z\pm 1$) and $s_i^x$ is the $x$ component of the spin-$\frac{1}{2}$ operator at a lattice site $i$. The transverse field $\Omega$ is assumed to be distributed regularly on the lattice sites while the local external longitudinal magnetic field $H_i$ on a lattice site $i$, is distributed via a given probability distribution. The exchange interaction $J_{ij}$ between the spins on the sites $i$ and $j$ takes the values according to the positions of the nearest neighbor spins. The two surfaces of the film have the intralayer coupling $J_1$. The interlayer coupling between the surface and its adjacent layer (i.e. layers $1,2$ and $L-1,L$) is denoted by $J_2$. For the rest of the layers, the interlayer and the intralayer couplings are assumed as $J_3$. The first summation in Eq. \re{denk1} is over the nearest-neighbor pairs of spins, and the remaining summations are over all the lattice sites. Magnetic fields are distributed on the lattice sites according to a given Gaussian probability distribution
\eq{denk2}{
P\paran{H_i}=\paran{\frac{1}{2\pi\sigma^2}}^{1/2}\exp{\paran{-\frac{H_i^2}{2\sigma^2}}}
} where $\sigma$ is the width of the zero magnetic field centered Gaussian distribution.

Number of $L$ different representative longitudinal magnetizations in $z$ direction ($m_i=\sandd{s_i^z},i=1,2,\ldots,L$) for the system can be given by usual EFT equations which are obtained by differential operator technique and decoupling approximation (DA) \cite{ref80,ref81},
\eq{denk3}{\begin{array}{lcl}
m_1&=&\left[A_1+m_1B_1\right]^z\left[A_2+m_2B_2\right]\\
m_L&=&\left[A_1+m_LB_1\right]^z\left[A_2+m_{L-1}B_2\right]\\
m_2&=&\left[A_3+m_2B_3\right]^z\left[A_2+m_1B_2\right]\left[A_3+m_3B_3\right]\\
m_{L-1}&=&\left[A_3+m_{L-1}B_3\right]^z\left[A_2+m_LB_2\right]\left[A_3+m_{L-2}B_3\right]\\
m_{k}&=&\left[A_3+m_{k}B_3\right]^z\left[A_3+m_{k-1}B_3\right]\left[A_3+m_{k+1}B_3\right],k=3,4,\ldots,L-2.\\
\end{array}}
Here $m_i,(i=1,2,\ldots, z)$ denotes the longitudinal magnetization of the $i^{th}$ layer. The coefficients in the expanded form of Eq. \re{denk3}
are given by \eq{denk4}{ A^k_p A^l_q B^m_p
B^n_q=\integ{}{}{}dH_iP\paran{H_i}
\cosh^k\paran{J_{p}\nabla}\cosh^l\paran{J_{q}\nabla}\sinh^m\paran{J_{p}\nabla}\sinh^n\paran{J_{q}\nabla}f\paran{H_i,x}|_{x=0}
} where $\nabla$ is the usual differential operator in the
differential operator technique  with indices $p,q=1,2,3$. The function $f\paran{H_i,x}$ is defined by
\eq{denk5}{f\paran{H_i,x}=\frac{x+H_i}{y}\tanh\paran{\beta y}, \quad y=\left[\paran{x+H_i}^2+\Omega^2\right]^{1/2}} as
usual for the spin-$\frac{1}{2}$ system with transverse  field. In Eq. \re{denk5}, $\beta=1/(k_B T)$ where $k_B$ is Boltzmann
constant and $T$ is the temperature. The effect of the exponential
differential operator to an arbitrary  function $F(x)$ is given by
\eq{denk6}{\exp{\paran{a\nabla}}F\paran{x}=F\paran{x+a}} with any
constant  $a$. DA will give the results of the Zernike approximation
\cite{ref82} for this system. Transverse magnetization for the $i^{th}$ layer ($m_i^{x}=\sandd{s_i^x},i=1,2,\ldots,L$) can be obtained by using the function
\eq{denk7}{f_x\paran{H_i,x}=\frac{\Omega}{y}\tanh\paran{\beta y}, \quad y=\left[\paran{x+H_i}^2+\Omega^2\right]^{1/2}}
in Eq. \re{denk4}  (and then in Eq. \re{denk3}) instead of Eq. \re{denk5}.

With the help of the Binomial expansion, Eq. \re{denk3} can be written in the form
\eq{denk8}{\begin{array}{lcl}
m_1&=&\summ{i=0}{z}{}\summ{j=0}{1}{}K_1\paran{i,j}m_1^i m_2^j\\
m_L&=&\summ{i=0}{z}{}\summ{j=0}{1}{}K_1\paran{i,j}m_{L}^i m_{L-1}^j\\
m_2&=&\summ{i=0}{z}{}\summ{j=0}{1}{}\summ{k=0}{1}{}K_2\paran{i,j,k}m_2^i m_1^j m_3^k \\
m_{L-1}&=&\summ{i=0}{z}{}\summ{j=0}{1}{}\summ{k=0}{1}{}K_2\paran{i,j,k}m_{L-1}^i m_{L}^j m_{L-2}^k \\
m_{k}&=&\summ{i=0}{z}{}\summ{j=0}{1}{}\summ{k=0}{1}{}K_3\paran{i,j,k}m_{k}^i m_{k-1}^j m_{k+1}^k \\
\end{array}} where

\eq{denk9}{\begin{array}{lcl}
K_1\paran{i,j}&=&\komb{z}{i}A_1^{z-i}A_2^{1-j}B_1^{i}B_2^{j}\\ 
K_2\paran{i,j,k}&=&\komb{z}{i}A_2^{1-j}A_3^{z+1-i-k}B_2^{j}B_3^{i+k}\\ 
K_3\paran{i,j,k}&=&\komb{z}{i}A_3^{z+2-i-j-k}B_3^{i+j+k}.\\ 
\end{array}}
These coefficients can be calculated from the definitions given in Eq. \re{denk4} with using Eqs. \re{denk5} and \re{denk6}.

For a given Hamiltonian and field distribution parameters, by determining the coefficients  from Eq. \re{denk9} we can obtain a system of coupled non linear equations from Eq. \re{denk8}, and by solving this system we can get the longitudinal magnetizations of each layer ($m_i,i=1,2,\ldots,L$). The longitudinal magnetization ($m$) and transverse magnetization ($m^{x}$)of the system can be calculated via
\eq{denk10}{m=\frac{1}{L}\summ{i=1}{L}{m_i}, \quad m^{x}=\frac{1}{L}\summ{i=1}{L}{m_i^{x}}.} Transverse  magnetization of each layer ($m_i^{x}$) can be obtained by constructing non linear equation system given in Eq. \re{denk8} by calculating coefficients from Eqs. \re{denk4} and \re{denk9}  using Eq. \re{denk7}, instead of Eq. \re{denk5}.

Since all longitudinal magnetizations are close to zero in the vicinity of the critical point, we can obtain another coupled  equation system to determine the transition temperature by linearizing the equation system given in  Eq. \re{denk8}, i.e.
\eq{denk11}{\begin{array}{lcl}
m_1&=&K_1\paran{1,0}m_1+K_1\paran{0,1}m_2\\
m_L&=&K_1\paran{1,0}m_{L}+K_1\paran{0,1}m_{L-1}\\
m_2&=&K_2\paran{1,0,0}m_2 +K_2\paran{0,1,0}m_1 +K_2\paran{0,0,1}m_3 \\
m_{L-1}&=&K_2\paran{1,0,0}m_{L-1} +K_2\paran{0,1,0}m_{L}  +K_2\paran{0,0,1}m_{L-2} \\
m_{k}&=&K_3\paran{1,0,0}m_{k}+K_3\paran{0,1,0}m_{k-1}+ K_3\paran{0,0,1} m_{k+1}.\\
\end{array}}

Critical temperature, as well as  critical transverse field can be determined from $\mathbf{\mathrm{det(A)=0}}$ where $A$ is the matrix of coefficients of the linear equation system given in Eq. \re{denk11}.

\section{Results and Discussion}\label{results}

In this section we discuss the effect of the Gaussian random magnetic field distribution on the phase diagrams of the system. We restrict ourselves in the case that all interactions are ferromagnetic i.e. $J_i>0, (i=1,2,3)$ and $z=4$ i.e. all layers in thin film have square lattices. We use the scaled interactions as
\eq{denk14}{J_3=J, \quad r_n=\frac{J_n}{J}, \quad n=1,2 \quad r_1=1+\Delta_s} Note that, the parameter $\Delta_s$ mostly used in literature in order to investigate the system according to
a surface exchange interaction in relation with other exchange interactions in the system. Since we want to concentrate on the effect of $\sigma$ and $\Delta_s$ on the phase diagrams and magnetizations, we choose $r_2=1.0$ in this work.

\subsection{Phase Diagrams}
Gaussian distribution defined in Eq.\re{denk2} is governed by only one parameter $\sigma$ , which is the
width of the distribution. This distribution distributes negative
and positive valued magnetic fields -which are chosen from the
zero centered Gaussian distribution- to lattice sites, so that the overall magnetic field is equal to zero. Although the total magnetic
field is zero, randomly distributed negative and positive fields
drag the system to the disordered phase. On the other hand, the
interactions $J_n,(n=1,2,3)$  enforce the system to stay in the
ordered phase. Another factor is the transverse field which forces the spins to align perpendicular to $z$ direction.
The last factor, the temperature which causes
thermal agitations induces a disordered phase when the energy
supplied by the temperature to the system is high enough. Thus a
competition takes place between these factors.

The effect of the Gaussian magnetic field distribution on the phase diagrams of the bulk systems is now well known. As $\sigma$ rises,  the
critical temperature decreases continuously which originates from the rising randomness of the orientations of the $s_i^z$. Similarly, rising $\Omega$ forces the spins to align perpendicular to the $z$ direction, and consequently it decreases the
critical temperature. Transition temperature reduces to zero at the critical value of transverse field $\Omega=\Omega_c$. After this value of $\Omega_c$, the system cannot exhibit an ordered phase, due to the lack of the chance of the spins to front in $z$ direction.

On the other hand, the phase diagrams for the Ising model with thin film geometry in the $(k_BT_c/J,\Delta_s)$ and $(\Omega_c/J,\Delta_s)$ plane are also well known (see e.g. Ref. \cite{ref49}). As $\Delta_s$ rises, the critical temperature and the critical transverse field of the system increases for arbitrary $L$. The curves for the different film thickness  ($L$) intersect at a special point which can be denoted by $(\Delta_s^{*},k_BT_c^{*}/J)$ and by $(\Delta_s^{*},\Omega_c^{*}/J$) for the curves in the ($k_BT_c/J,\Delta_s$) and ($\Omega_c/J,\Delta_s$) planes, respectively. For the values of $\Delta_s>\Delta_s^{*}$, thin films have higher critical temperature (and critical transverse field) than the bulk system while the reverse is valid for $\Delta_s<\Delta_s^{*}$. For the value of $\Delta_s=\Delta_s^{*}$, thin film with thickness $L$ has the same critical values of the bulk, independent of the $L$. The value of $\Delta_s$ also changes  the relation between the film thickness and  the critical temperature (and critical transverse field) of the film. The thicker films have higher critical values for $\Delta_s<\Delta_s^{*}$ while thicker films have lower critical values for $\Delta_s>\Delta_s^{*}$ than thinner ones.

After this short summary, at first let us investigate the effect of the Gaussian random field distribution width on the phase diagrams of the transverse Ising thin film in $(k_BT_c/J,\Delta_s)$ and $(\Omega_c/J,\Delta_s)$ planes. In Figs.  \re{sek1} and \re{sek2}, phase diagrams can be seen for different $L$ values  with chosen $\sigma$. We can see from Figs. \re{sek1} and \re{sek2} that, rising $\sigma$ decreases the critical temperatures and critical transverse fields as in the bulk counterpart of discussed system. This decrement in the values of the $k_BT_c/J$ and $\Omega_c/J$ shows itself also in the special point coordinates $k_BT_c^{*}/J$ and $\Omega_c^{*}/J$ while there is no significant change in the value of the $\Delta_s^{*}$ when $\sigma$ rises. We note that for $\sigma=0.0$, $\Delta_s^{*}$ value that makes critical temperature at $\Omega=0.0$ independent of the film thickness ($L$) is $\Delta_s^{*}=0.3068$ and the value of $\Delta_s^{*}$  that makes critical transverse field independent of the film thickness ($L$) is $\Delta_s^{*}=0.3326$ as in Ref. \cite{ref49}. Also, rising $\sigma$ does not affect the relation between the film thickness and critical temperature (or transverse field) , i.e. for $\Delta_s<\Delta_s^{*}$ thicker films have higher critical values and reverse is true for $\Delta_s>\Delta_s^{*}$ with $\sigma\ne 0$, if the special point is present. When $\sigma$ is large enough, special point disappears and thicker films have lower critical values for a certain $\Delta_s$ as seen in Figs. \re{sek1}(f) and \re{sek2}(f).

The variation of the $k_BT_c^{*}/J$ and $\Omega_c^{*}/J$ with $\sigma$ can be seen in Fig. \re{sek3}. This variation is nothing but the variation of the
$k_BT_c/J$ (at $\Omega=0$) and $\Omega_c/J$ with $\sigma$ for the TIM in the bulk system, since the special point is defined as the intersection point of the curves in the $(k_BT_c/J,\Delta_s)$ (and $(\Omega_c/J,\Delta_s)$) plane for different film thicknesses and with the bulk curve in the same plane, which is just a parallel line to the $\Delta_s$ axis. We note that the curves in Fig. \re{sek3} ends at the point $\sigma=4.22$, i.e. for the values that provide $\sigma>4.22$, the system can not exhibit special point.

\begin{figure}[h]\begin{center}
\epsfig{file=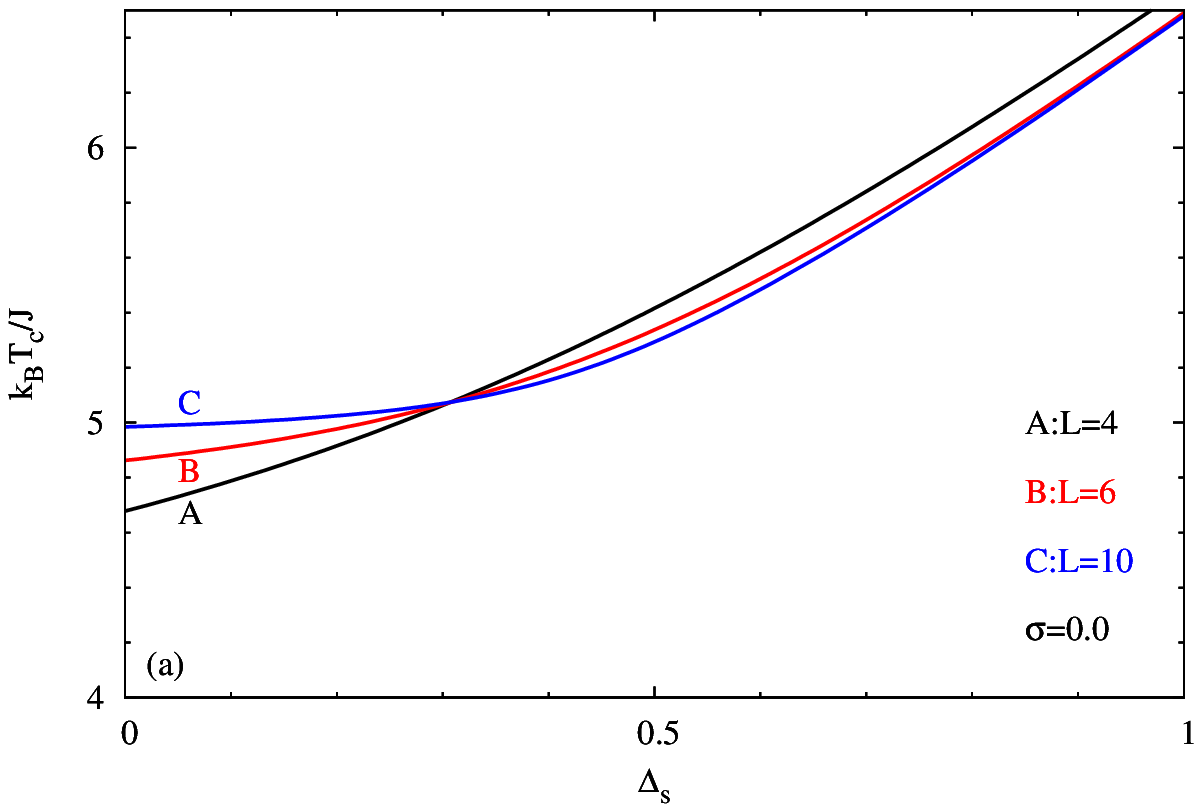, width=6cm}
\epsfig{file=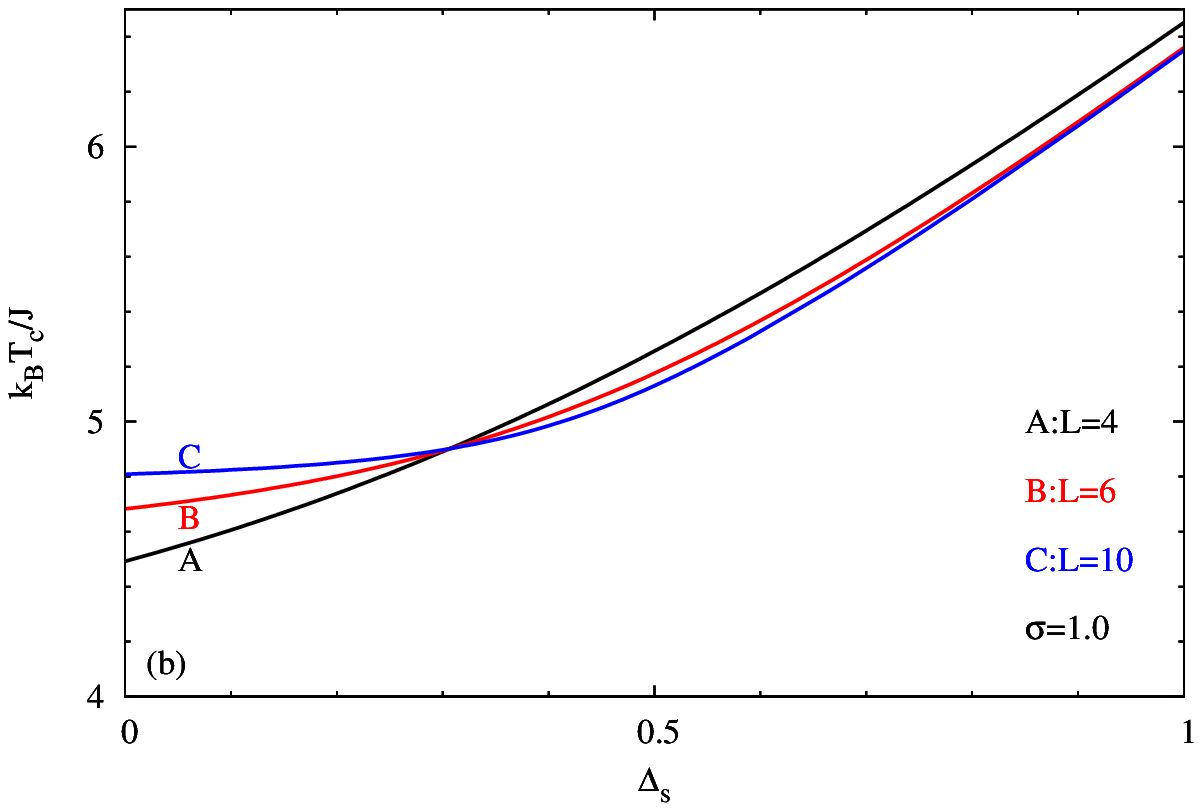, width=6cm}
\epsfig{file=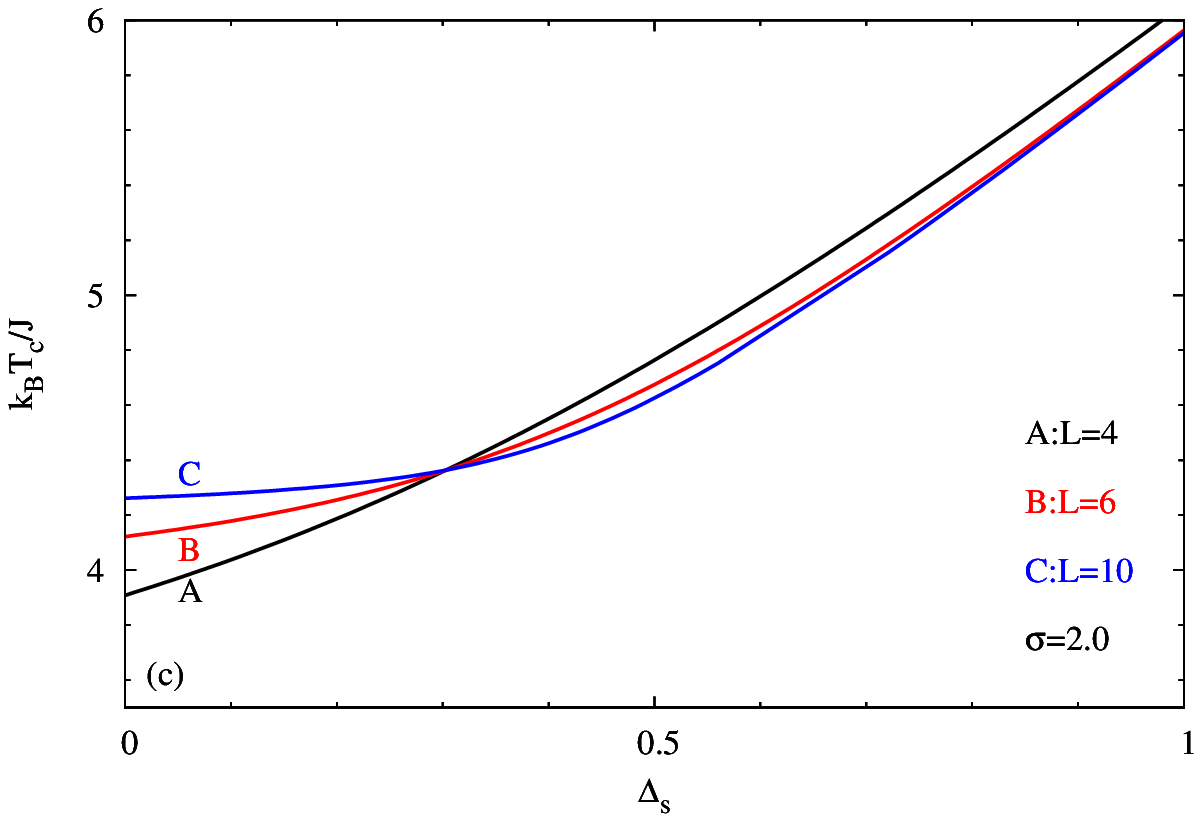, width=6cm}
\epsfig{file=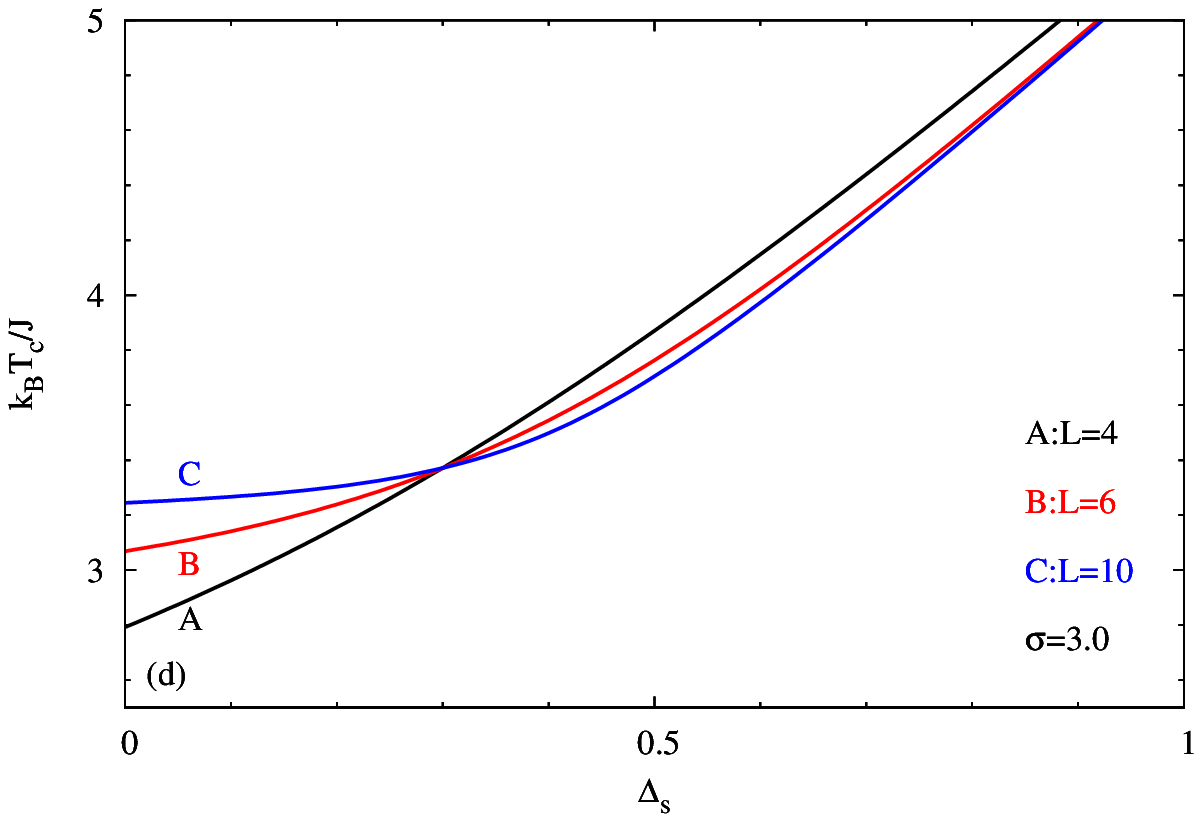, width=6cm}
\epsfig{file=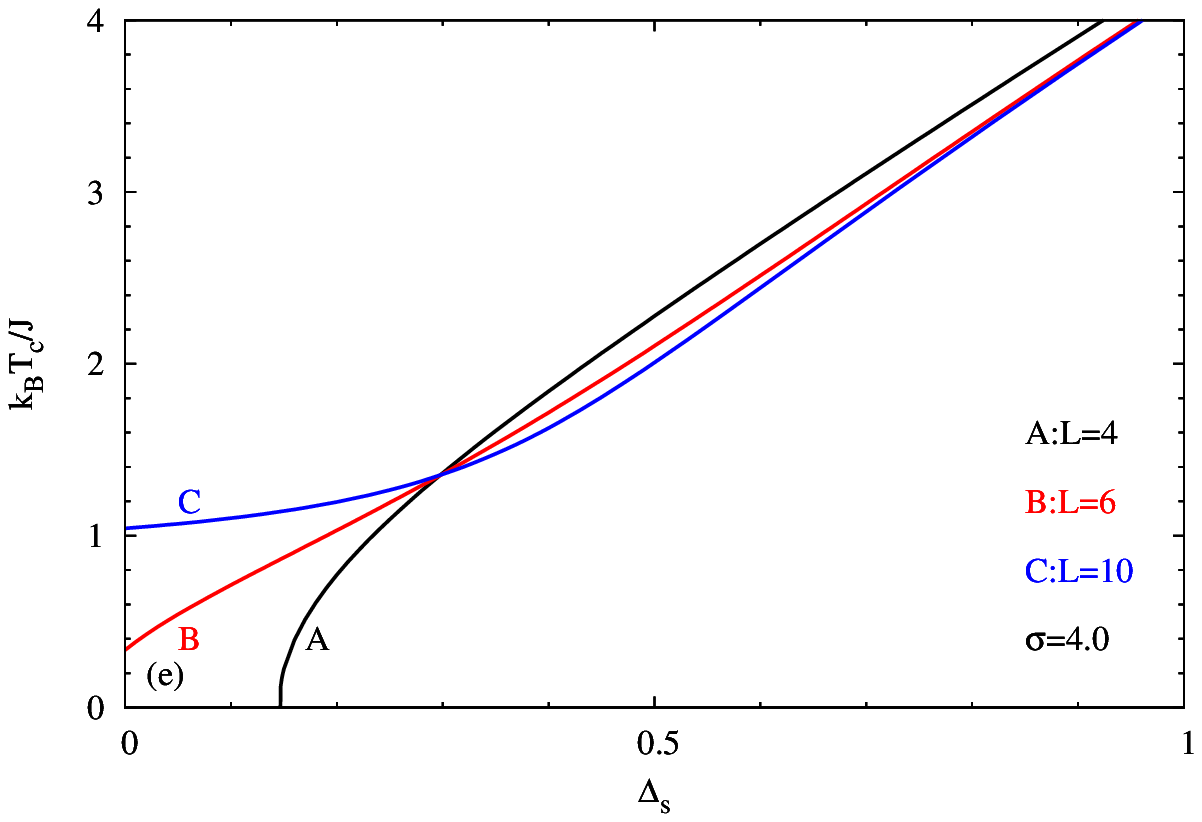, width=6cm}
\epsfig{file=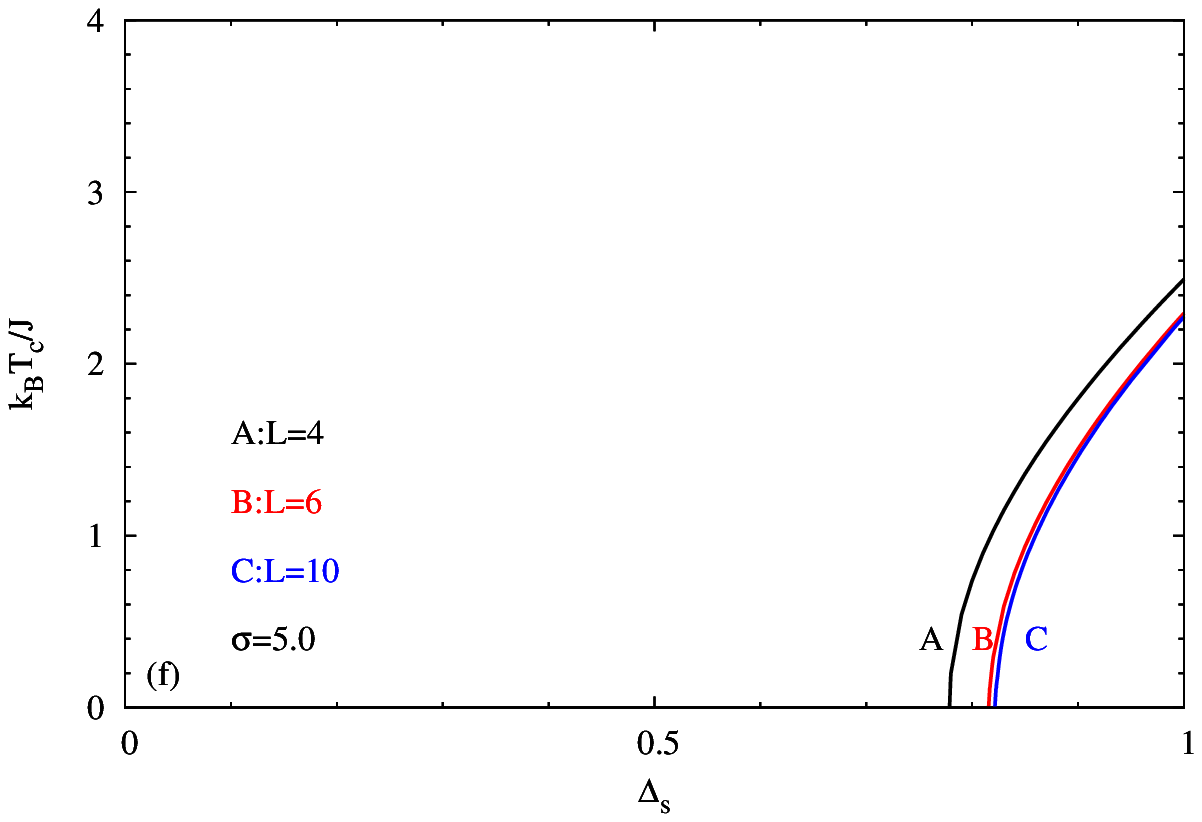, width=6cm}
\end{center}
\caption{The phase diagrams of the thin film described by the TIM  with single Gaussian random
field distribution in  the ($k_BT_c/J,\Delta_s$) plane for different $\sigma$ values with selected thickness values of $L=4,6,10$ . The selected value of $r_2$ is $r_2=1$
} \label{sek1}\end{figure}

\begin{figure}[h]\begin{center}
\epsfig{file=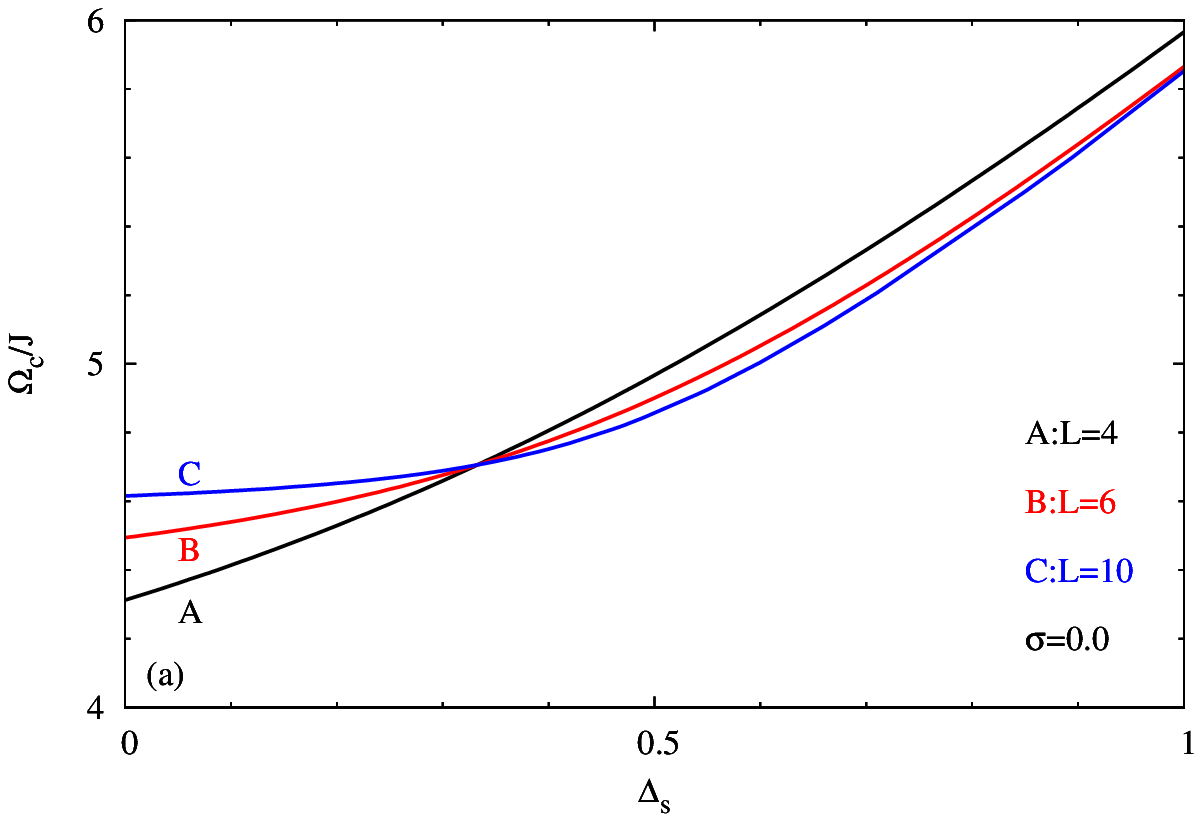, width=6cm}
\epsfig{file=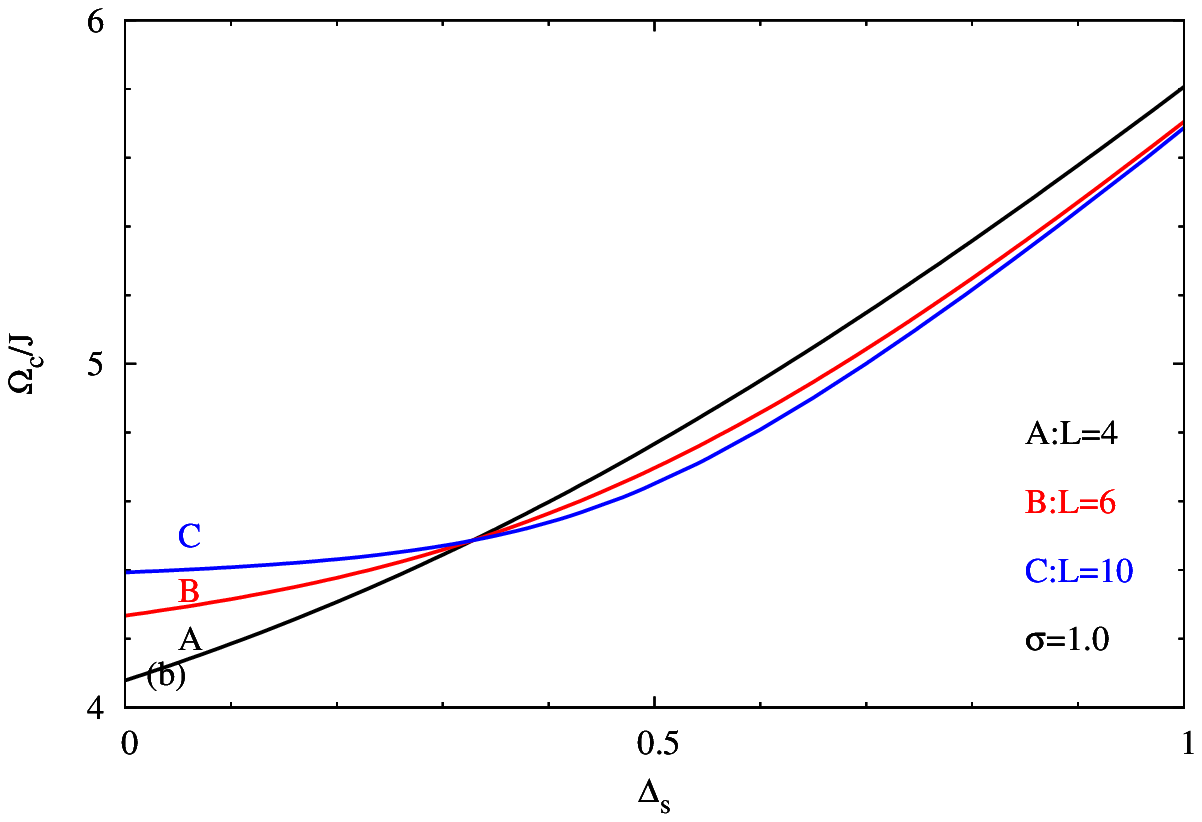, width=6cm}
\epsfig{file=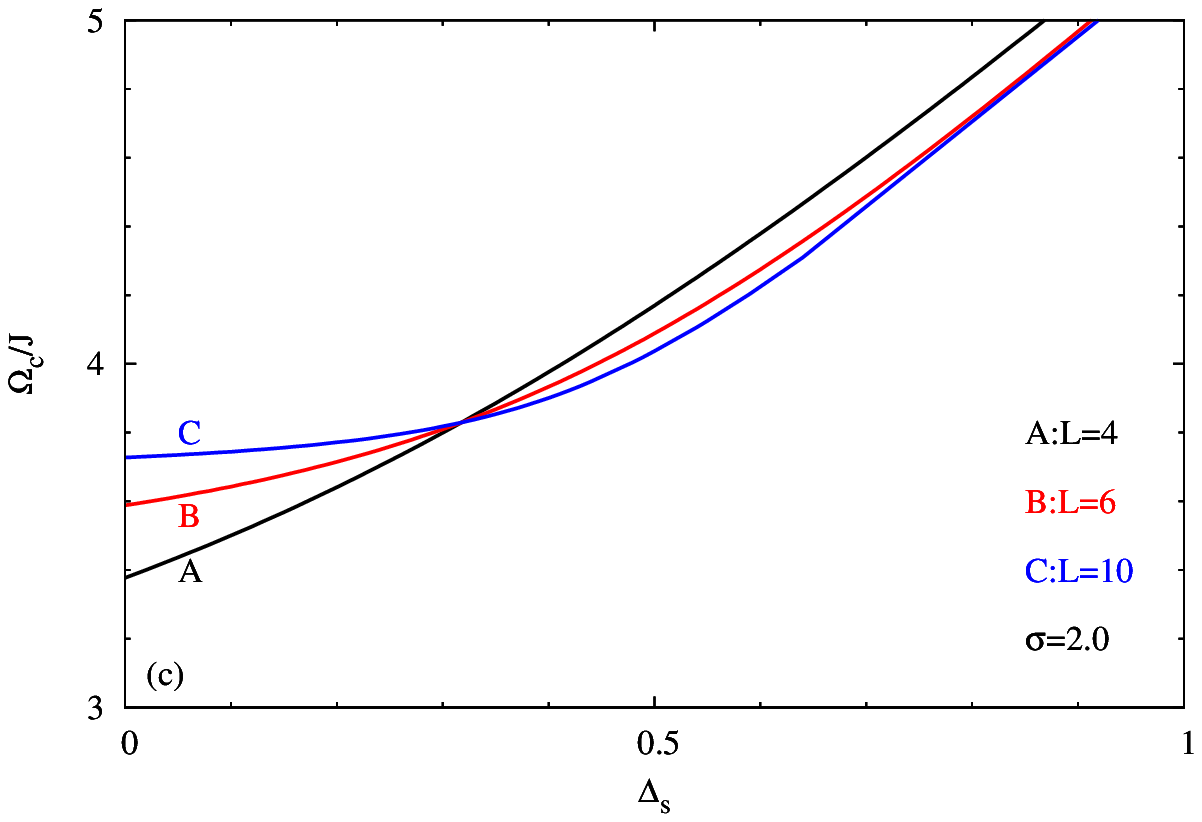, width=6cm}
\epsfig{file=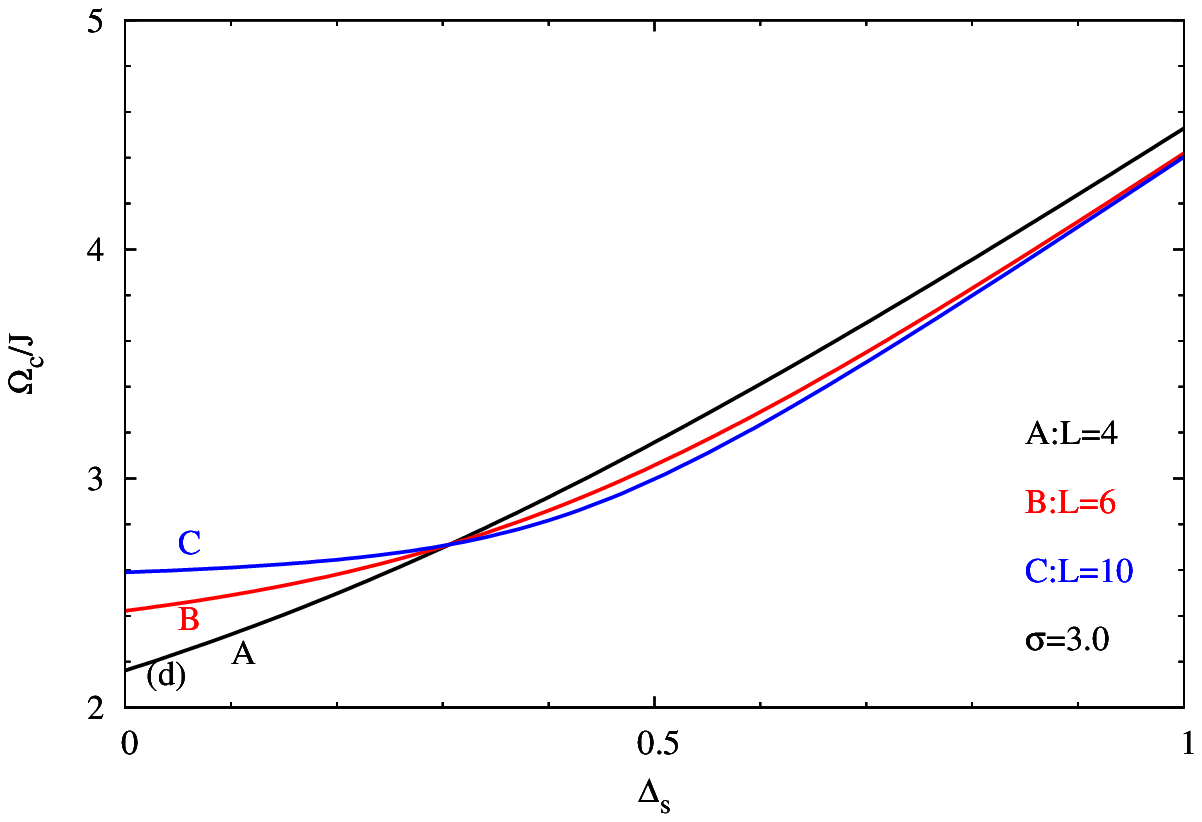, width=6cm}
\epsfig{file=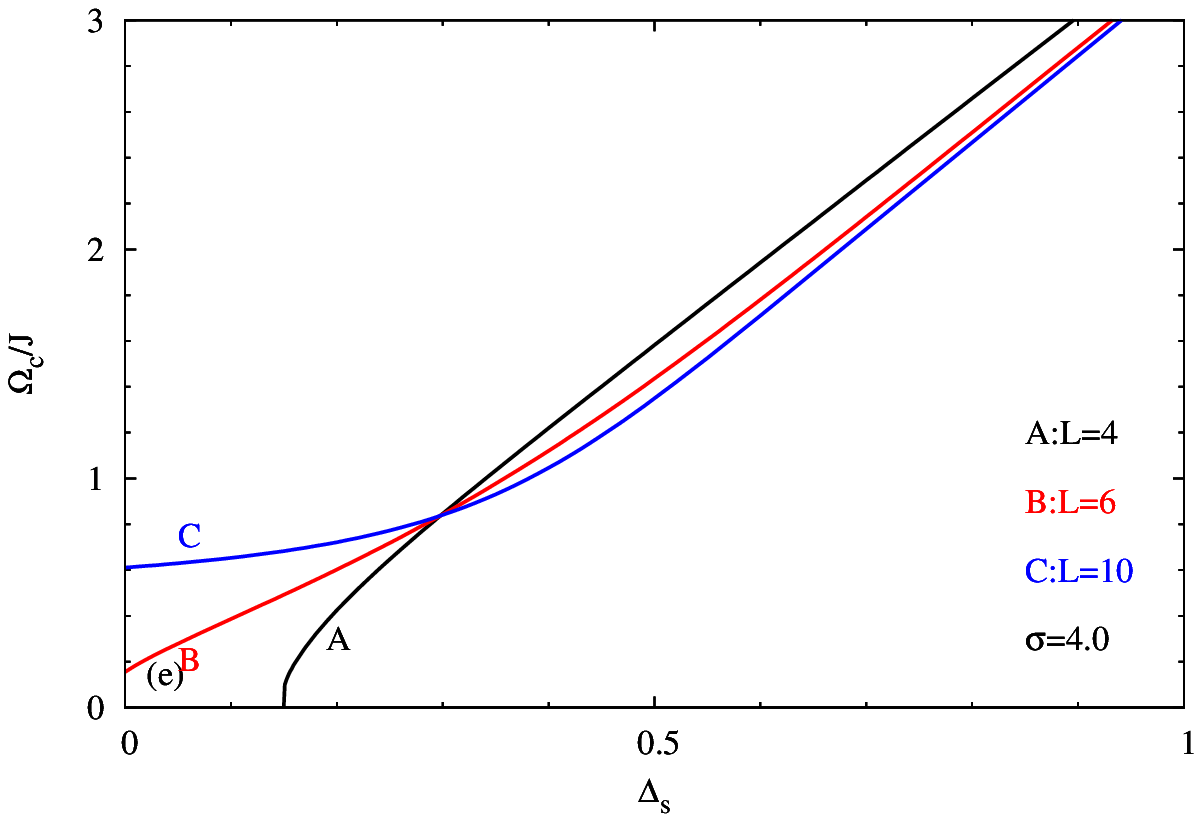, width=6cm}
\epsfig{file=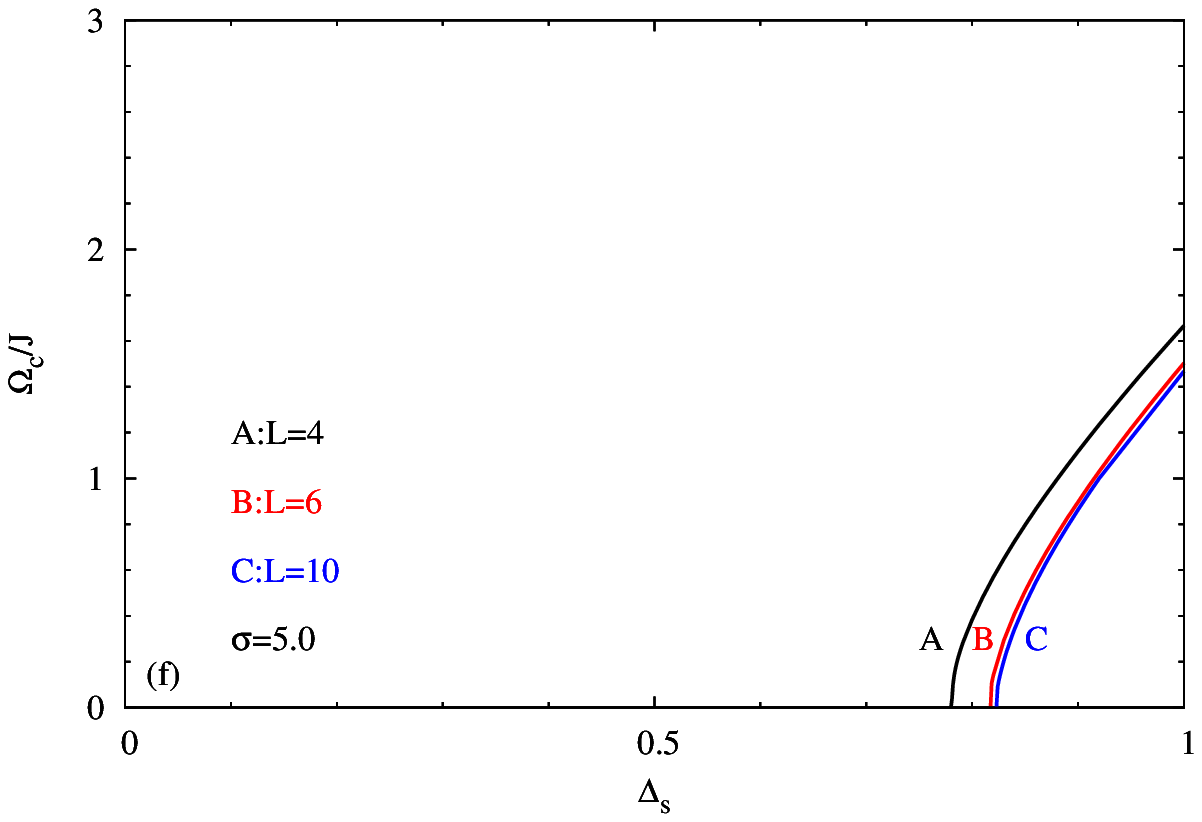, width=6cm}
\end{center}
\caption{The phase diagrams of the thin film described by the TIM  with single Gaussian random
field distribution in  the ($\Omega_c/J,\Delta_s$) plane for different $\sigma$ values with selected thickness values of $L=4,6,10$ . The selected value of $r_2$ is $r_2=1$
} \label{sek2}\end{figure}

\begin{figure}[h]\begin{center}
\epsfig{file=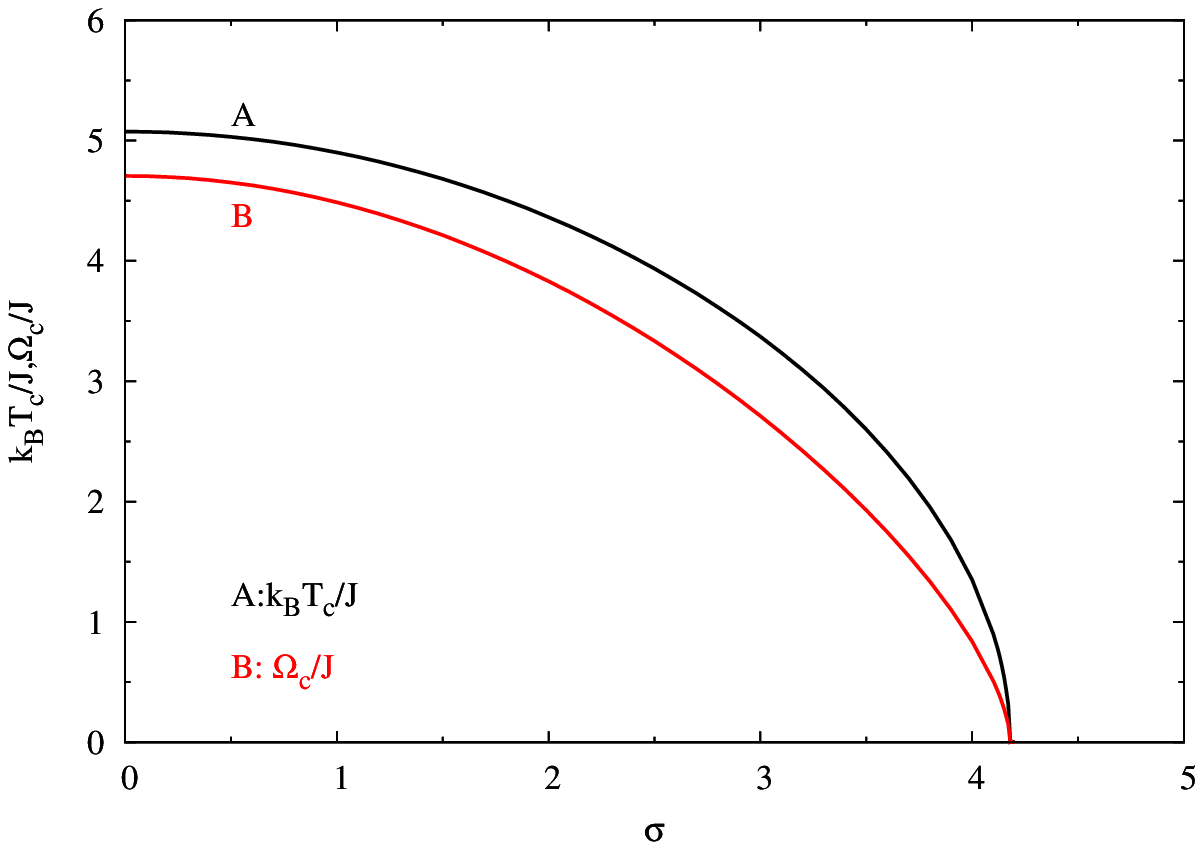, width=6cm}
\end{center}
\caption{Variation of $k_BT_c^{*}/J$ and $\Omega_c^{*}/J$ with $\sigma$ for the thin film described by the TIM  with single Gaussian random
field distribution.} \label{sek3}\end{figure}

The phase diagrams in  $(k_BT_c/J,\Omega/J)$ plane can be seen for the values of $\Delta_s=0.0$ and $\Delta_s=0.6$ (which provide the conditions $\Delta_s<\Delta_s^{*}$ and $\Delta_s>\Delta_s^{*}$ respectively) with film thicknesses $L=4,6,10$ and for different $\sigma$ values in Fig. \re{sek4}. We can see from  Fig. \re{sek4} that the qualitative relation between critical temperatures (and  critical transverse fields) and film thickness is also valid for these diagrams; i.e. for  $\Delta_s=0.0<\Delta_s^{*}$, all critical temperatures increase whereas for  $\Delta_s=0.6>\Delta_s^{*}$, all critical temperatures decrease when the film thickness increases for given $\Delta_s$ and any $\sigma$ . As in the  bulk systems, rising $\sigma$ shrinks the ferromagnetic region in the $(k_BT_c/J,\Omega/J)$ plane. This fact is due to two reasons: rising $\Omega$ forces the spins to align perpendicular to the $z$ direction and rising $\sigma$ forces the spins to align parallel to $z$ direction, but randomly i.e. $+z$ or $-z$. Two of these results to the decline of the order parameter.  Now, in order to understand how $\sigma$ effects the phase transition beside the transverse field, let us look at the effect of these parameters on the longitudinal and transverse magnetization.

\begin{figure}[h]\begin{center}
\epsfig{file=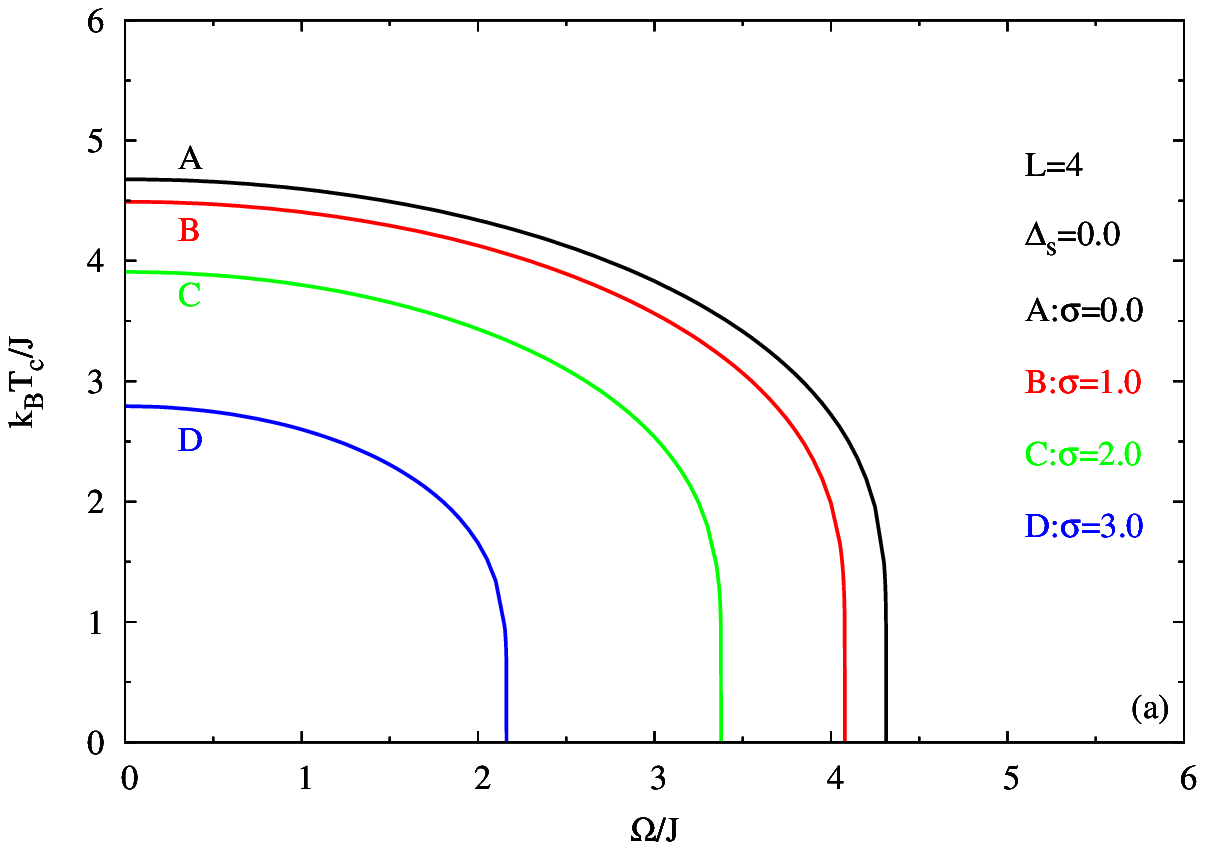, width=6cm}
\epsfig{file=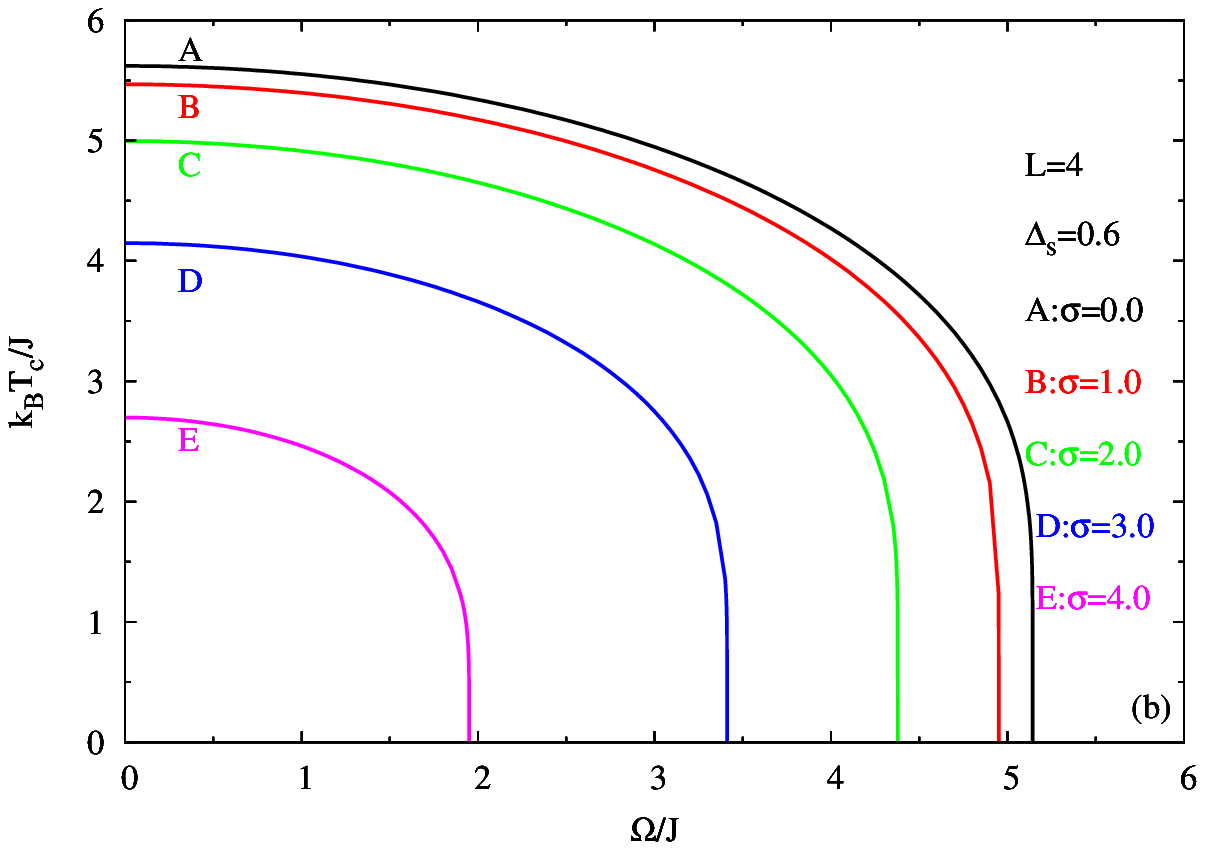, width=6cm}
\epsfig{file=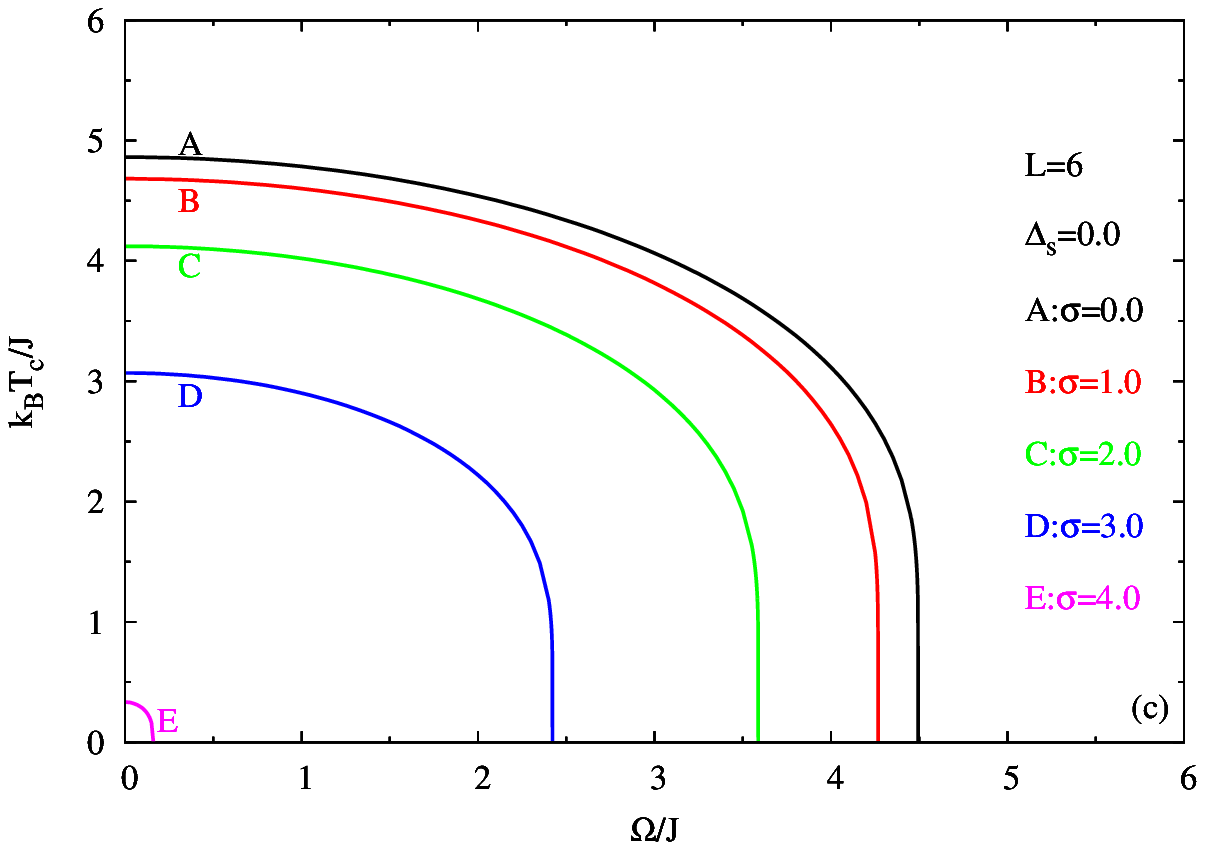, width=6cm}
\epsfig{file=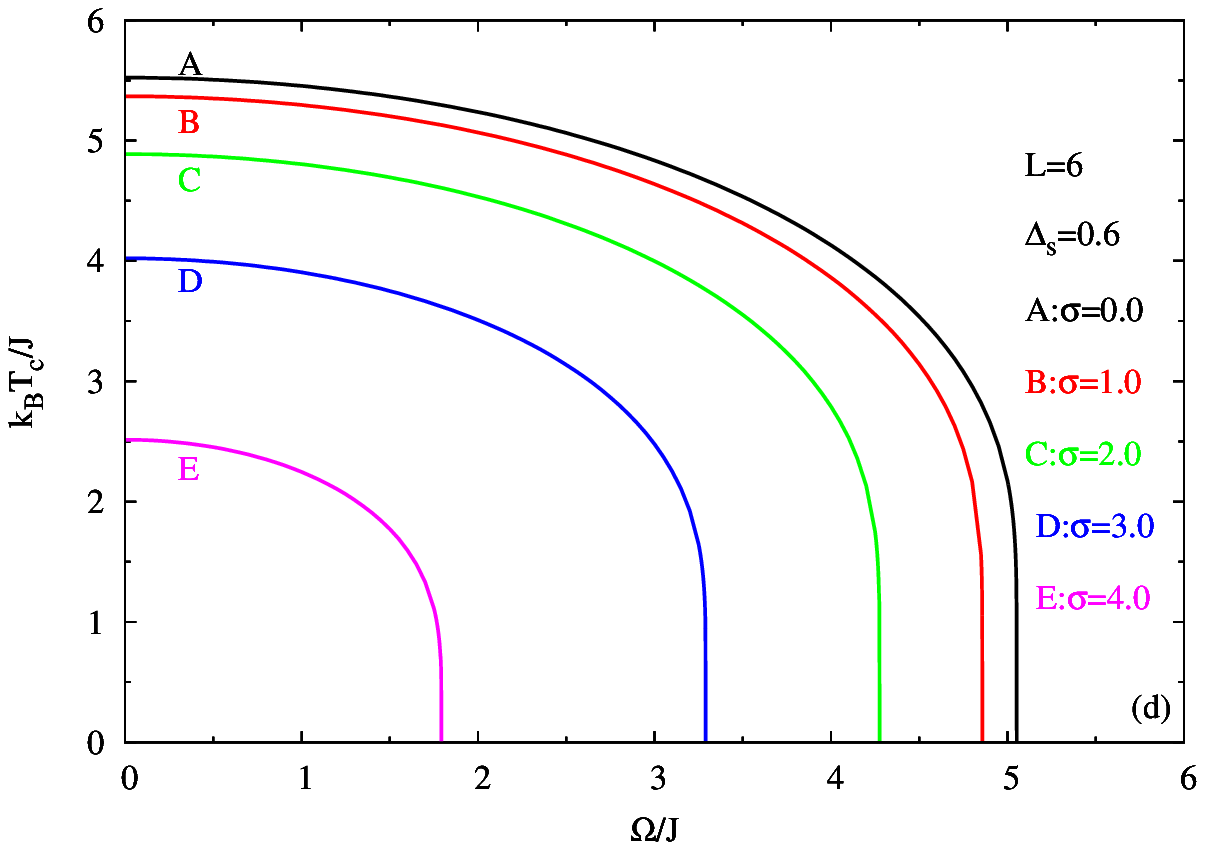, width=6cm}
\epsfig{file=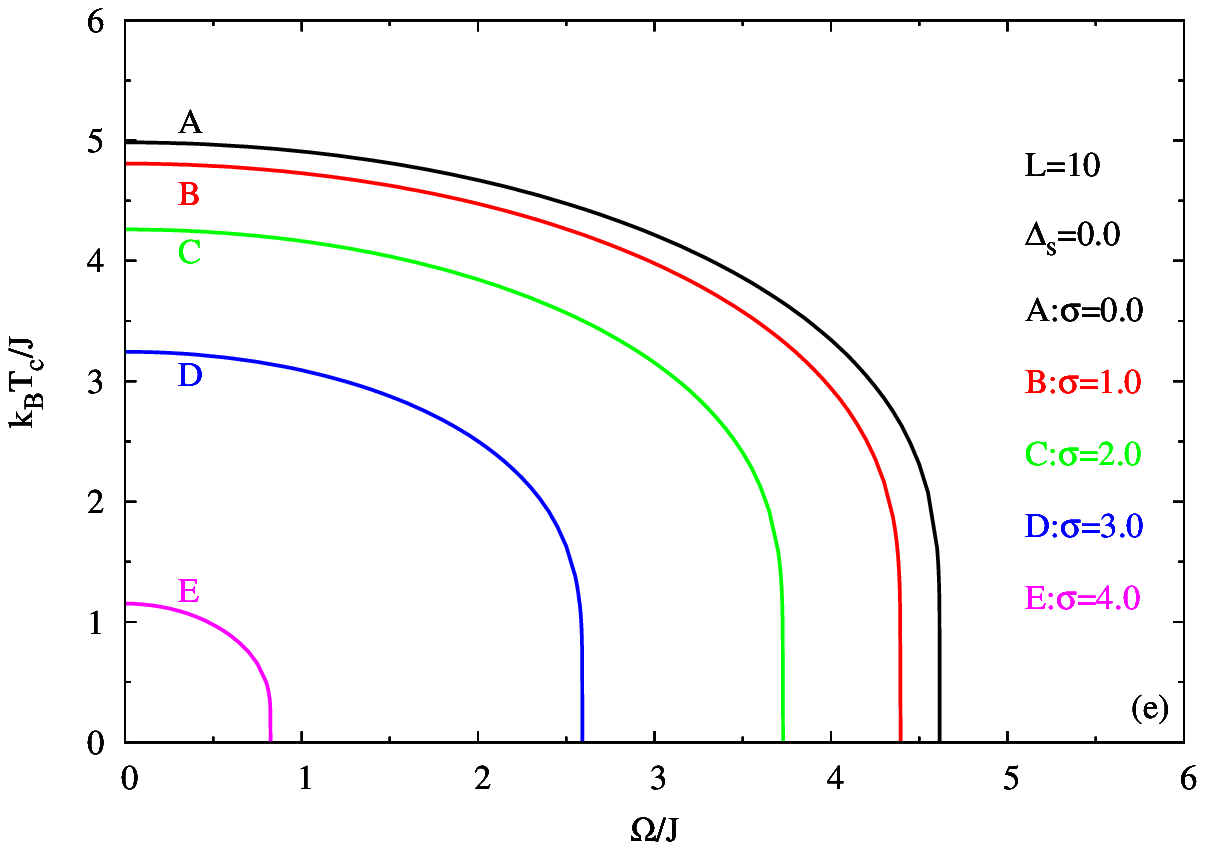, width=6cm}
\epsfig{file=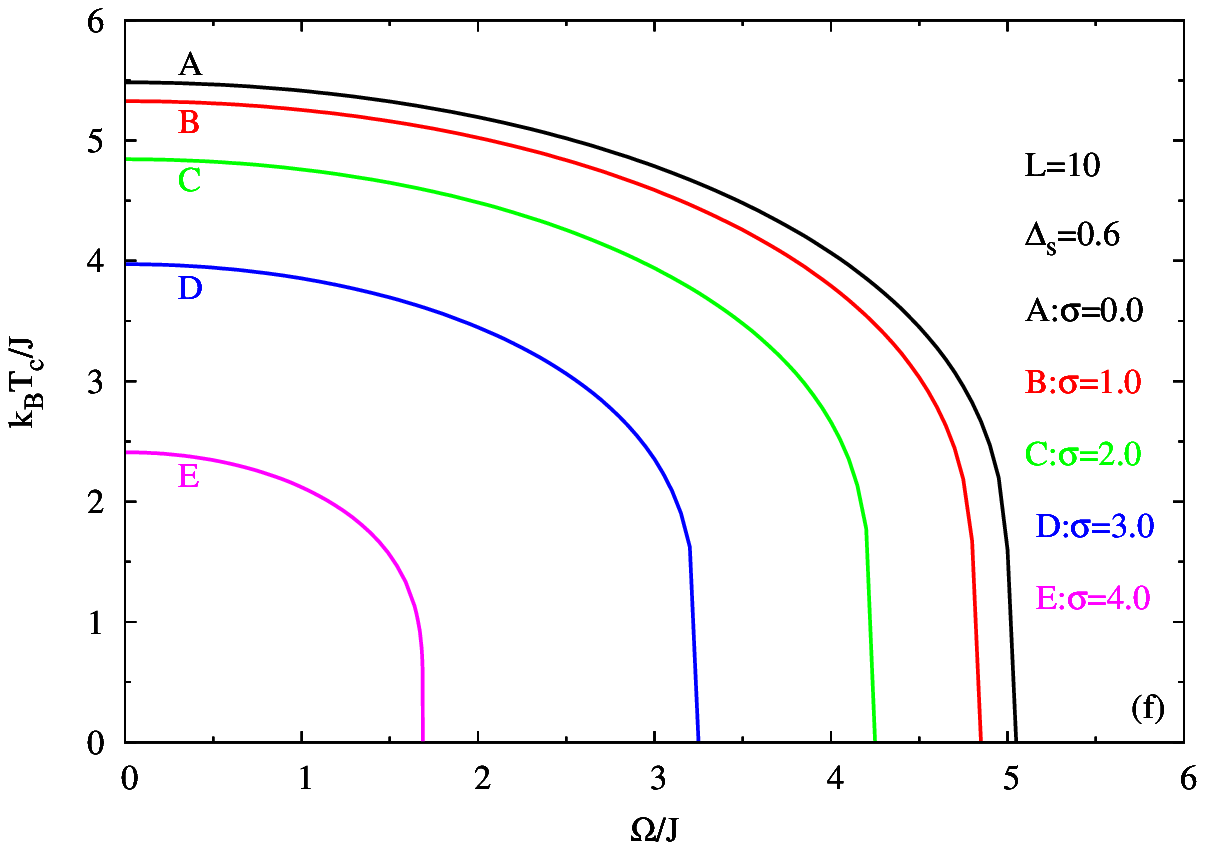, width=6cm}
\end{center}
\caption{The phase diagrams of the thin film described by the TIM  with single Gaussian random
field distribution in the ($k_BT_c/J,\Omega/J$) plane for different $\sigma$ values with selected thickness values of $L=4,6,10$ . The chosen value of $r_2$ is $r_2=1$
} \label{sek4}\end{figure}


\subsection{Magnetization Behaviors}

Let us investigate the effect of the random Gaussian magnetic field distribution on the magnetization behaviors with temperature for both the longitudinal and transverse counterparts. In order to  make inferences about how $\sigma$ effects the variation of longitudinal magnetization with temperature, we present the curves for the parameter values $L=4,10$, $\Omega=0.0,2.0$ and $\Delta_s=0.0,0.6$ through large set of parameters as representatives. These curves can be seen in Fig. \re{sek5} for the film thickness $L=4$, and in Fig. \re{sek6} for the film thickness $L=10$. Each plot contains surface longitudinal magnetization ($m_1$) (with solid lines) and  inner layer longitudinal magnetization ($m_{L/2}$) (with dotted lines), in order to investigate how this randomness spreads into the film.

Firstly, there is no need to elaborate on the general effect that, rising $\sigma$ decreases the longitudinal magnetization at any temperature that lower than the critical temperature.
At first glance, from Figs. \re{sek5}(a),(c) and \re{sek6} (a),(c), we can see that, in the absence of the transverse field, rising $\sigma$ separates the  ground state longitudinal magnetizations of the surface and inner layer, for a given $L$ and $\Delta_s$. The value of the difference between the ground state longitudinal magnetization of the surface and the inner layer depends on $\sigma$, $L$ and $\Delta_s$ in the absence of the transverse field. As we can see from Figs. \re{sek5}(b),(d) and \re{sek6} (b),(d), the same effect can be achieved by rising $\Omega$ even in the absence of random field distribution (see the curves labeled by A in Figs. \re{sek5} (b), (d) and \re{sek6} (b), (d)). Rising $\Omega$ results in that more spins align perpendicular to $z$ direction. In the case of $\Delta_s=0.0$, i.e. all exchange interactions are equal to  each other, since the spins which are located in the inner layers have one more neighbor spin than  the spins located at the surface layer; it is expected that the inner spins have more chance to align in $z$ direction than that of the surface spins. Accordingly, we get higher longitudinal magnetization in the inner layers than that of the surface layer (the curve labeled by A in Figs. \re{sek5} (b) and \re{sek6} (b)).
Rising of $\Delta_s$  compensates the lack of a neighbor site of the surface spins, then after some value of $\Delta_s$, the surface layer magnetization becomes greater than  that of the inner ones (the curve labeled A in Figs. \re{sek5} (d) and \re{sek6} (d)). The same scenario is also valid for the random field case. Rising $\sigma$ causes the random alignment of all spins in $\pm z$ direction in the system. Because of the existence of an additional spin interaction, the spins which are located in the inner layers have more chance to align parallel to each other than the spins in the surface layer. When $\Delta_s$ increases, after some value of it, this situation turns reverse in the presence of Gaussian random field distribution (for instance, compare the curves labeled C of Figs. \re{sek5} (a) and \re{sek5} (c)).

Increasing $\sigma$ causes an increment in the difference between the longitudinal magnetization of the surface  layer and the inner layer at any temperature (that little than the critical temperature) and transverse field value (see for example the curves labeled C and D in Fig. \re{sek5}(c)). As $L$ increases,  this difference also increases for large enough $\sigma$ values (see for example the curves labeled D in Fig. \re{sek5}(c) and \re{sek6}(c)), and rising $\Delta_s$ value at first reduces this difference and after a specific value at which the longitudinal magnetization of the surface and inner layer become equal to each other, rising $\Delta_s$ increases the difference between the two magnetization profiles, and the longitudinal magnetization of the surface becomes greater than that of the inner one.

In order to focus on the difference between the ground state longitudinal magnetization of the surface ($m_1$) and that of the inner layer ($m_{L/2}$), we present the variation of the $\Delta m =m_1-m_{L/2}$ curves with the transverse field at the temperature $k_BT_c/J=0.001$ for some selected values of $\Delta_s,L$ and $\sigma$ in Fig. \re{sek7}. At this temperature, the energy supplied from the temperature to the system is very small in comparison with the spin-spin interaction, then the situation may be treated as the investigation of ground state magnetization.  In the absence of the randomly distributed  field (i.e. $\sigma=0.0$), $\Delta m\ne 0$ value originates completely from the transverse field. The general trend of variation of $\Delta m$ with $\Omega$ can be seen in Figs. \re{sek7} (a), (e) in the absence of the magnetic field. Rising $\Omega$ can not create significant change until $\Omega/J=1.0$, then $\mutlak{\Delta m}$ increases with increasing $\Omega/J$ until the specific value of $\Omega/J$ for a given $\Delta_s$ and $L$. After then,  $\mutlak{\Delta m}$ approaches to zero and finally at the value of $\Omega_c/J$ for a given  $\Delta_s$ and $L$, $\mutlak{\Delta m}$ falls to zero because all layers exhibit zero magnetization. We can see from Fig. \re{sek7} that this trend is also valid in the presence of the randomly distributed longitudinal magnetic field. For instance, we can see from Fig. \re{sek7} (b) that the shapes of the curves are similar to that given in  Fig. \re{sek7} (a) but with lower $\Omega_c/J$ and higher value of $\mutlak{\Delta m}$ at any $\Omega/J$ (even in the absence of the transverse field) for fixed $\Delta_s$. This fact comes from the randomly distributed field which has mechanism explained above. But undoubtedly, the net effect of the transverse field and randomly distributed magnetic field on the $\Delta m$ is not simply the sum of them, since the random magnetic field distribution also affects the transverse magnetization which originates from the transverse field. We can make one more conclusion that, the higher $\sigma$ curves have slightly higher maximum value of the $\mutlak{\Delta m}$,  which can be seen in Figs. \re{sek7} (c) and (d) by comparing the curves labeled by A,B,C,D.

Finally, we want to make some conclusions on the effect of the $\sigma$ on the ground state transverse magnetization of the surface layer and an inner layer. For this purpose, let us investigate the situation for the values of $\Omega$ that provides the condition $\Omega/J>1$, since only these values makes $\Delta m$ different from zero in the absence of the magnetic field, as seen in Fig. \re{sek7} (a). We again focus on the $\Delta m^x=m_1^{x}-m_{L/2}^{x}$ calculated at temperature $k_BT_c/J=0.001$.  We can see the variation of the $m_1^{x}$ and $m_{L/2}^{x}$ with $\sigma$ for $L=4$ and $\Omega=2.0$ and $\Delta_s=0.0,0.4$ in Fig. \re{sek8} (a). We can see that, rising $\sigma$ decreases both $m_1^{x}$ and $m_{L/2}^{x}$ values, which can be explained by the fact that when $\sigma$ rises, more and more spins are forced to align in $\pm z$ direction and this causes a decline in the transverse magnetization. As discussed above, $\Delta m^x$ is positive at $\Delta_s=0$ while it turns to negative when $\Delta_s=0.4$ at any $\sigma$, which can be seen in Fig. \re{sek8} (a). If we look at the variation of the $m_{1}^{x}-\sigma$ and $m_{L/2}^{x}-\sigma$ curves calculated at a temperature $k_BT_c/J=0.001$ for $\Omega=1.3$ and $L=4$, we can see from Fig. \re{sek8} (b) that some $m_{1}^{x}$ curves (for the values $\Delta_s=0.14,0.16$) are located above the $m_{L/2}^{x}$ curve , and some of them  (for the values $\Delta_s=0.24,0.26$) are below the $m_{L/2}^{x}$ curve for all values of $\sigma$. But the curves of $m_{1}^{x}$ for the values  $\Delta_s=0.18,0.20,0.22$ intersects the curve $m_{(L/2)}^{x}$ while $\sigma$ rises. This means that for some values of $\Delta_s$, $\Delta m^x$  is always positive or negative  (i.e. for all $\sigma$), but for some values of $\Delta_s$,  $\Delta m^x$ changes its sign as $\sigma$ varies. These $\Delta_s$ values are of course depend on the other parameters of the system such as $L,\Omega$. The borders of the positive valued $\Delta m^x$ and negative valued $\Delta m^x$ (calculated at the temperature $k_BT_c/J=0.001$) can be seen in Fig. \re{sek9} for $L=4$ and for selected values of $\Delta_s=0.18,0.20,0.22,0.24$ in the $(\sigma,\Omega/J)$ plane. Ground state $\Delta m^x$ calculated at $\Delta_s$  is positive on the region that covers the right side of the related $\Delta_s$ curve in the $(\sigma,\Omega/J)$ plane and negative on the region that covers left side of the related $\Delta_s$ curve. We can see from the Fig. \re{sek9} that, rising $\Delta_s$ shifts the curves to the right and this means that expanding the region that $\Delta m^x<0$. The absence of $\Delta_s<0.18$ curves in Fig. \re{sek9} tells us that, for $\Delta_s<0.18$ always $\Delta m^x>0$ within the region in Fig. \re{sek9} and similarly, the absence of $\Delta_s>0.24$ curves in Fig. \re{sek9} means  that, for $\Delta_s>0.24$ always $\Delta m^x<0$ within the region in Fig. \re{sek9}. For a fixed value of $\sigma$, and for example with  $\Delta_s=0.22$, when $\Omega$ increases, the system passes from the region $\Delta m^x<0$ to the region $\Delta m^x>0$ at a specific value of $\Omega/J$. But, for instance for a fixed value of $\Omega/J=1.6$, rising $\sigma$ crosses from the region $\Delta m^x<0$ to the region $\Delta m^x>0$ for a certain $\sigma$, then with rising $\sigma$ crosses $\Delta m^x<0$ again, according to the curve related to the $\Delta_s=0.22$ (which is labeled as C in Fig. \re{sek9}). This shows us that, indeed a competition between the Gaussian distributed longitudinal magnetic field and the transverse field, results some complicated situations on the transverse magnetization.

\begin{figure}[h]\begin{center}
\epsfig{file=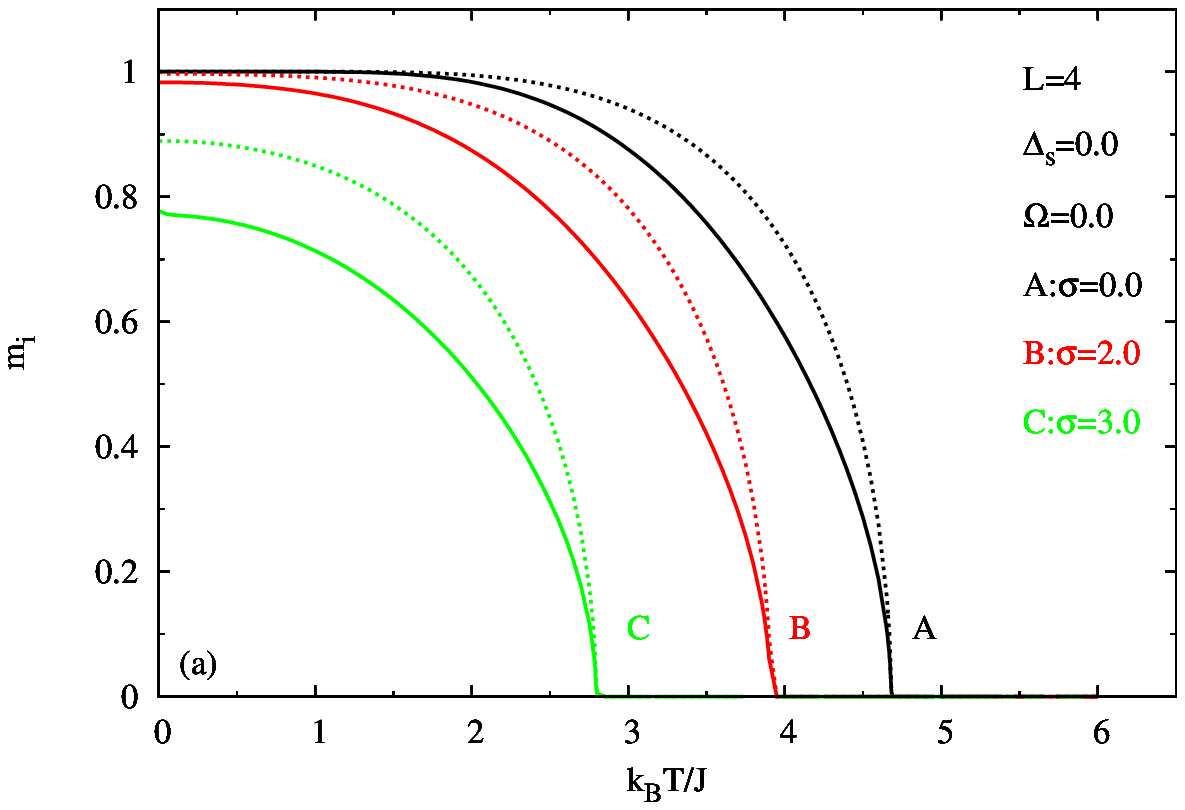, width=6cm}
\epsfig{file=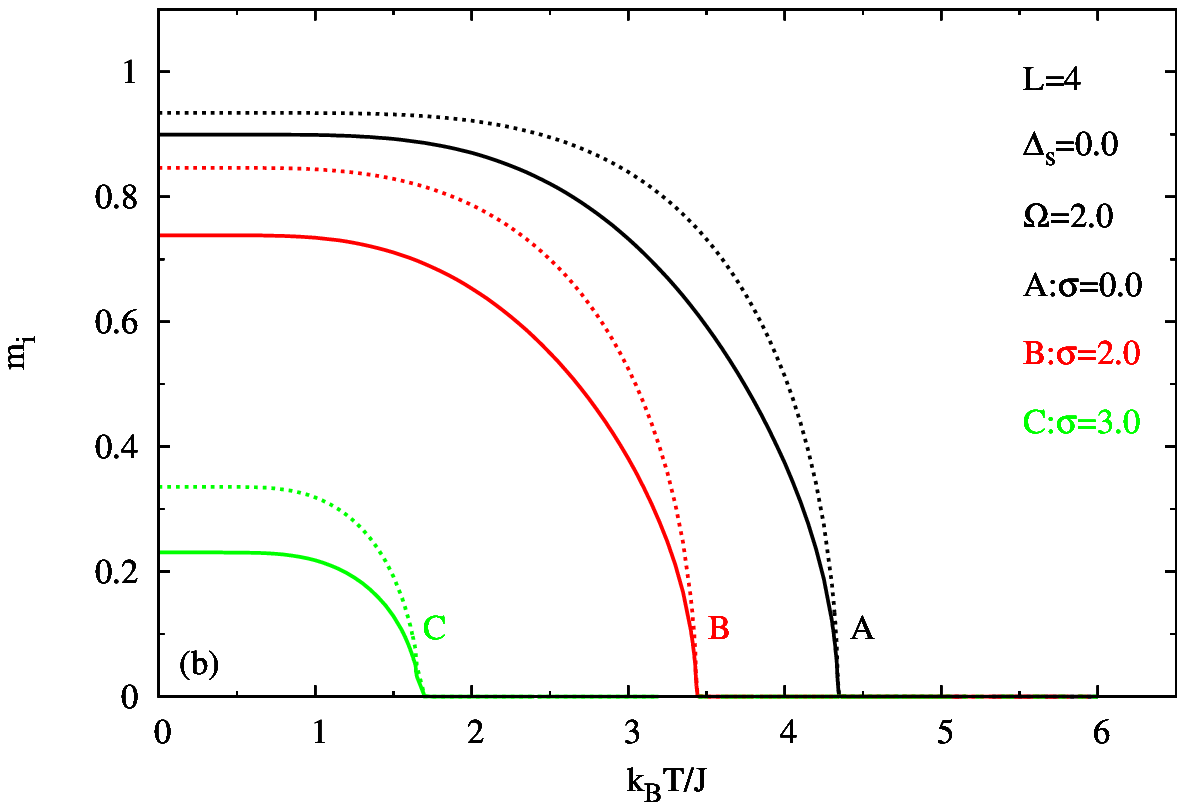, width=6cm}
\epsfig{file=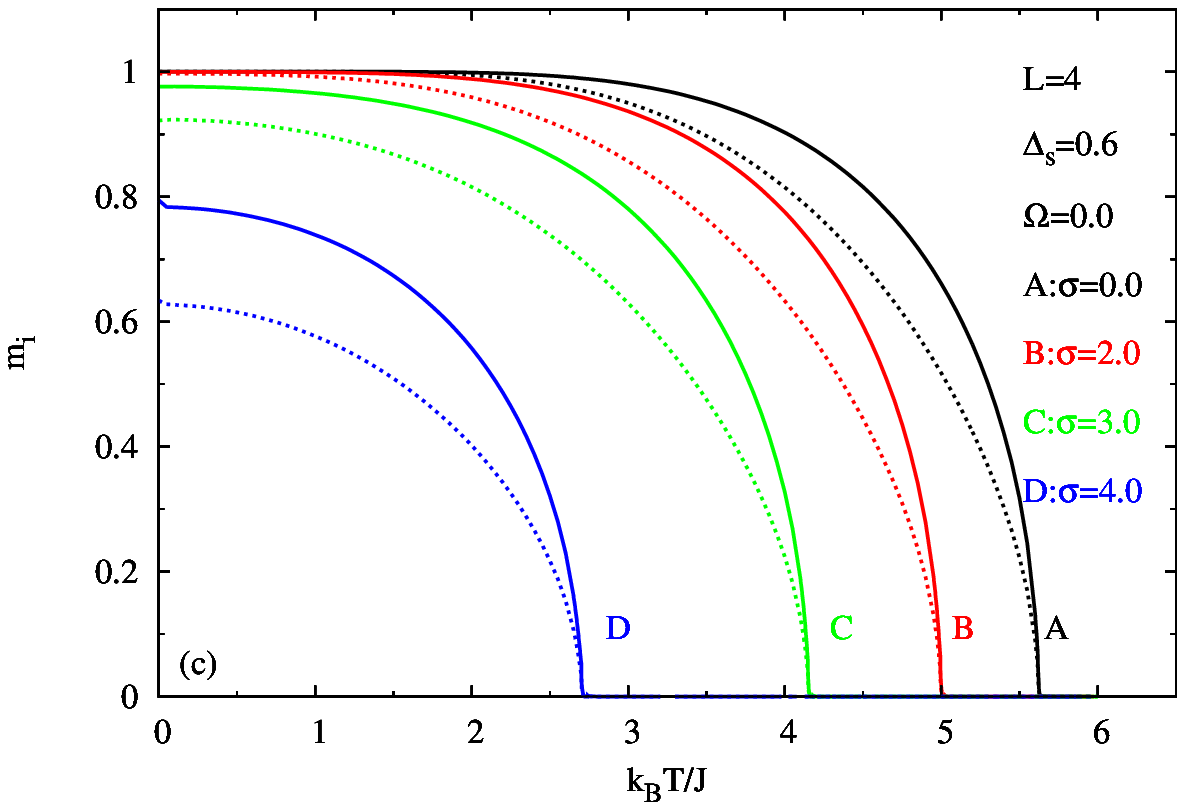, width=6cm}
\epsfig{file=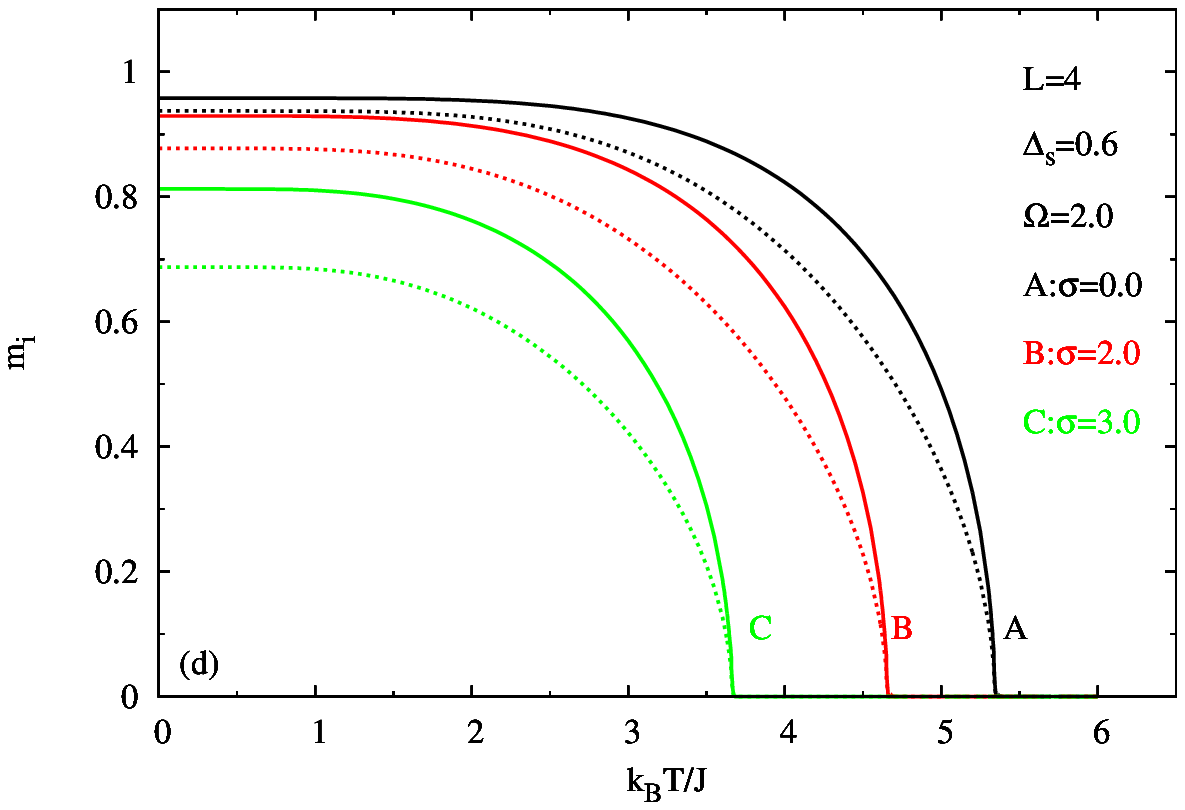, width=6cm}
\end{center}
\caption{Variation of the longitudinal layer magnetizations with temperature for some selected  values of $\Delta_s,\Omega$ and $\sigma$ for the thin film described by the TIM  with single Gaussian random field distribution with thickness $L=4$. Surface longitudinal magnetization $m_1$ is represented  by solid line while the longitudinal  magnetization of the $L/2$ indexed  layer ($m_{L/2}$) is represented by dotted line.  The chosen value of $r_2$ is $r_2=1$.}
\label{sek5}\end{figure}

\begin{figure}[h]\begin{center}
\epsfig{file=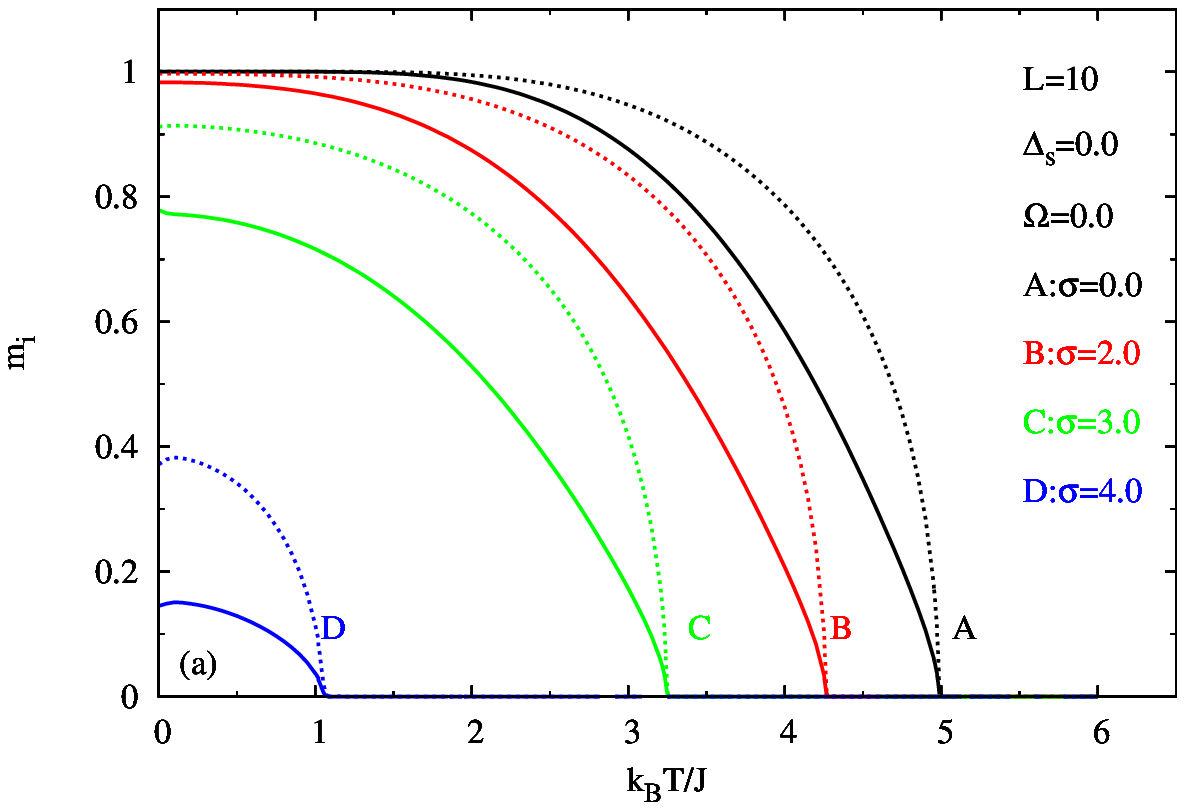, width=6cm}
\epsfig{file=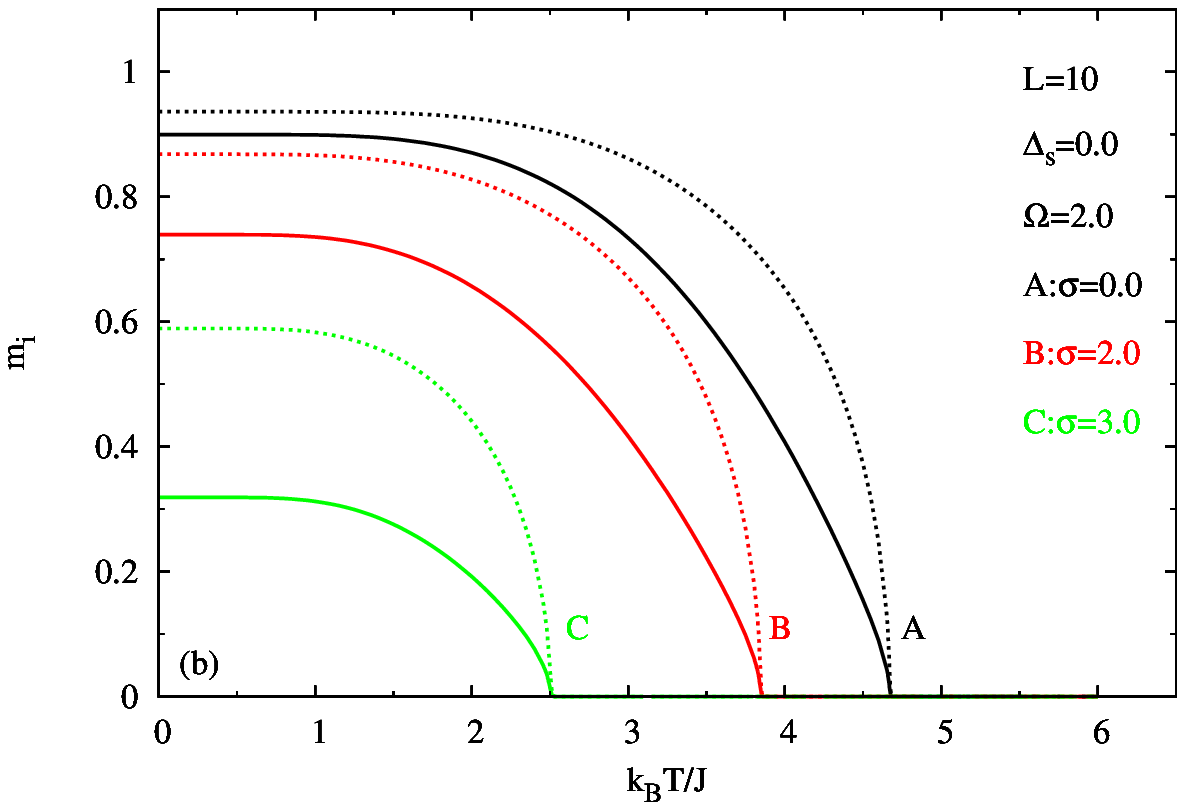, width=6cm}
\epsfig{file=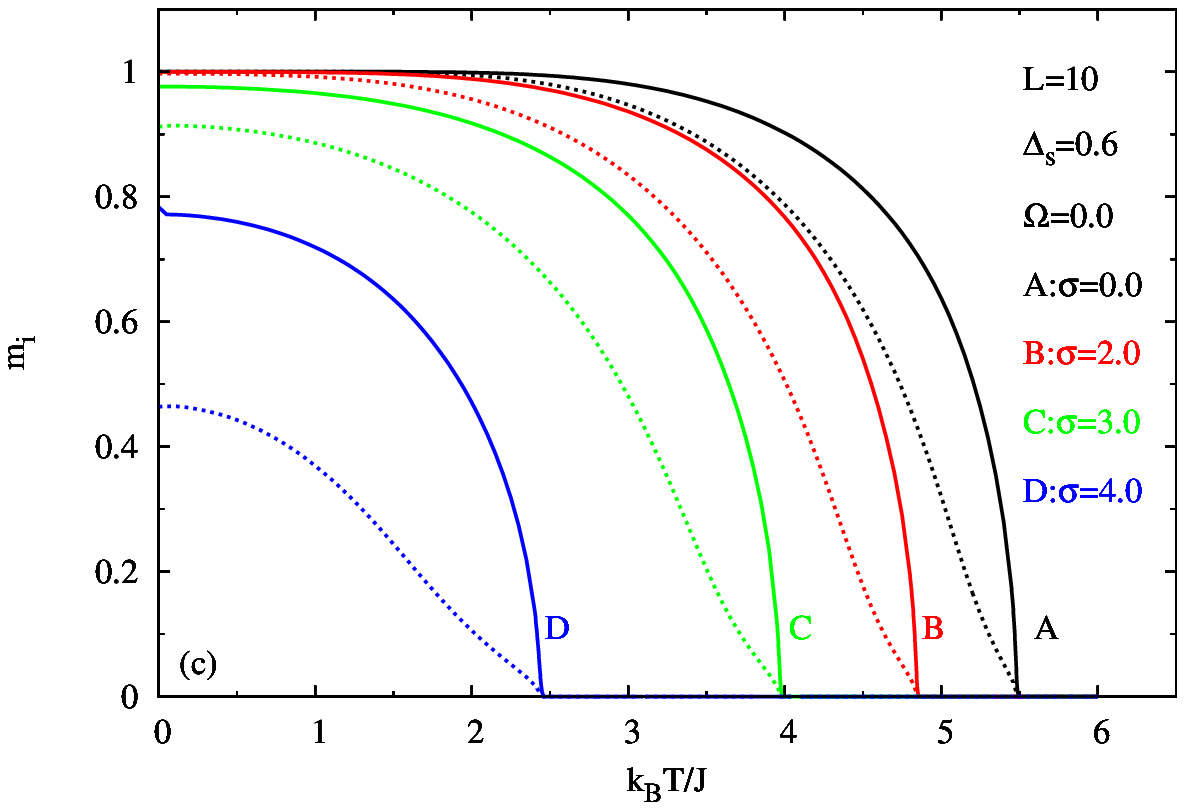, width=6cm}
\epsfig{file=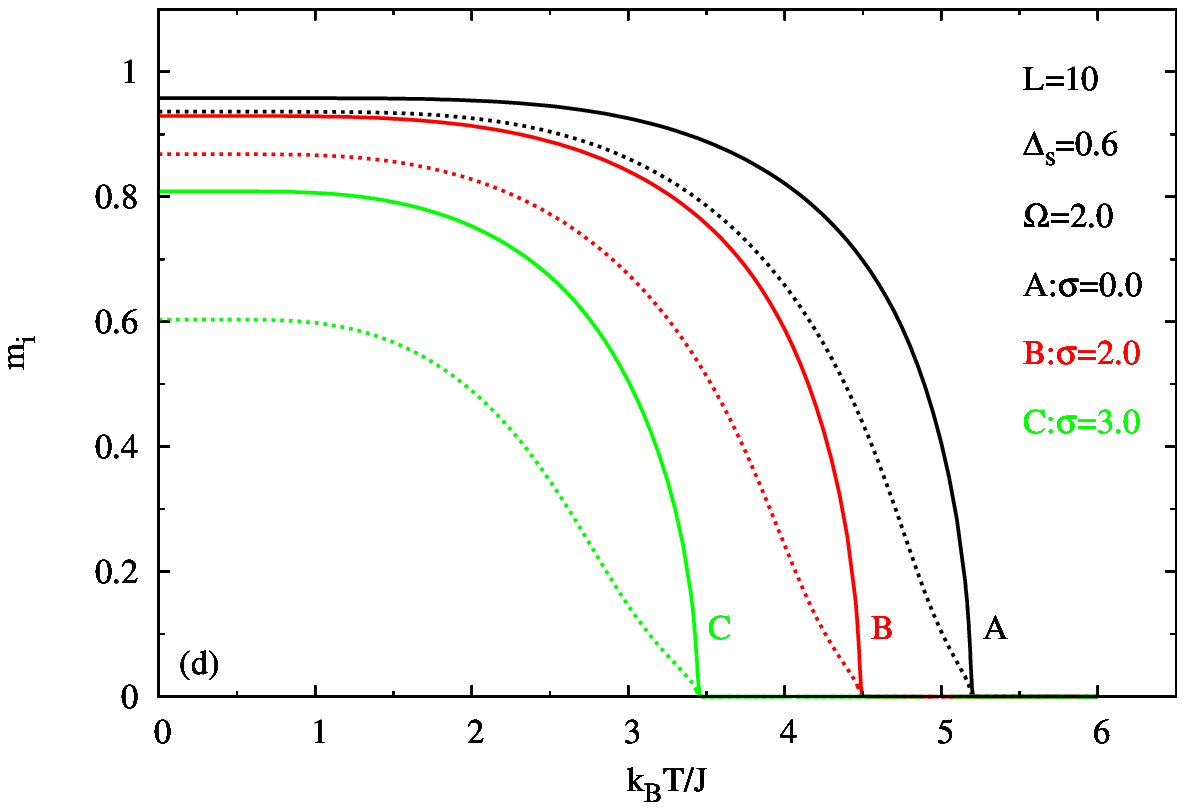, width=6cm}
\end{center}
\caption{Variation of the longitudinal layer magnetizations with temperature for some selected values of $\Delta_s,\Omega$ and $\sigma$ for the thin film described by the TIM  with single Gaussian random field distribution with thickness $L=10$. Surface longitudinal magnetization $m_1$ is represented by solid line while longitudinal  magnetization of the $L/2$ indexed  layer ($m_{L/2}$) is represented by dotted line. Chosen value of $r_2$ is $r_2=1$.}
\label{sek6}\end{figure}

\begin{figure}[h]\begin{center}
\epsfig{file=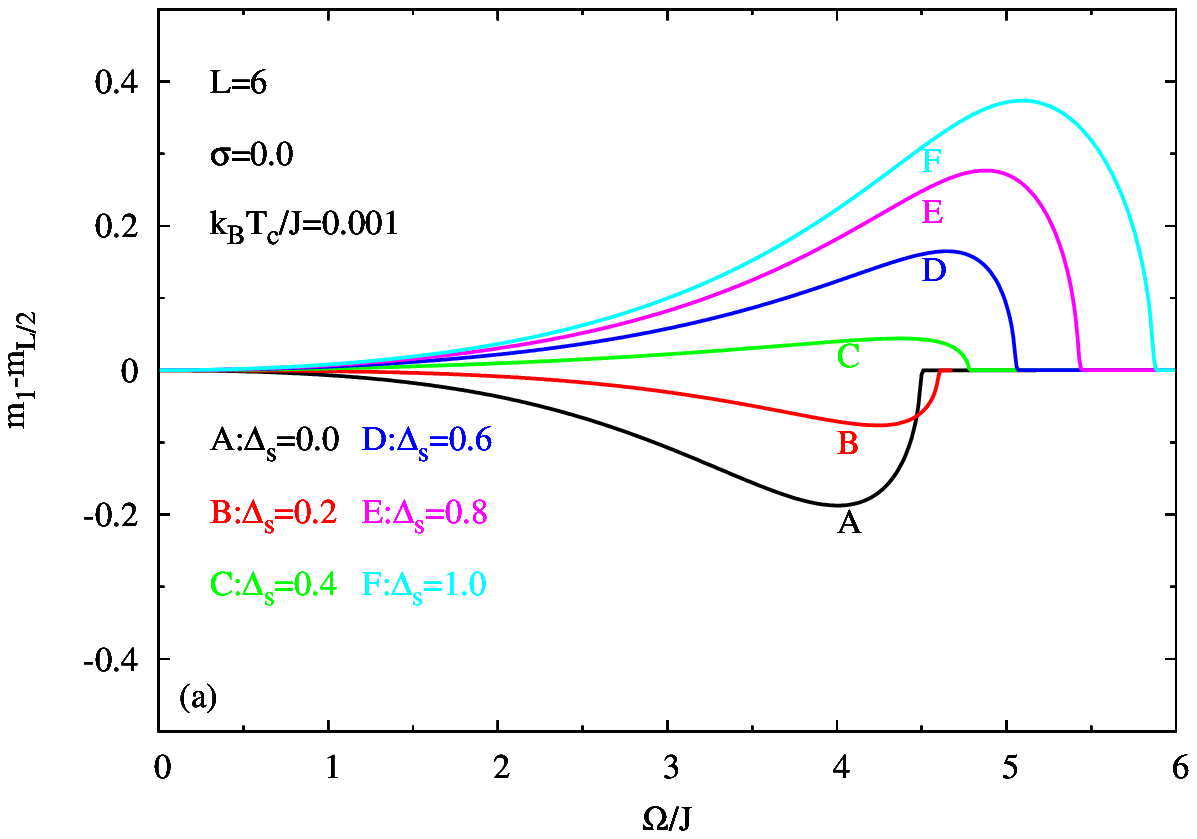, width=6cm}
\epsfig{file=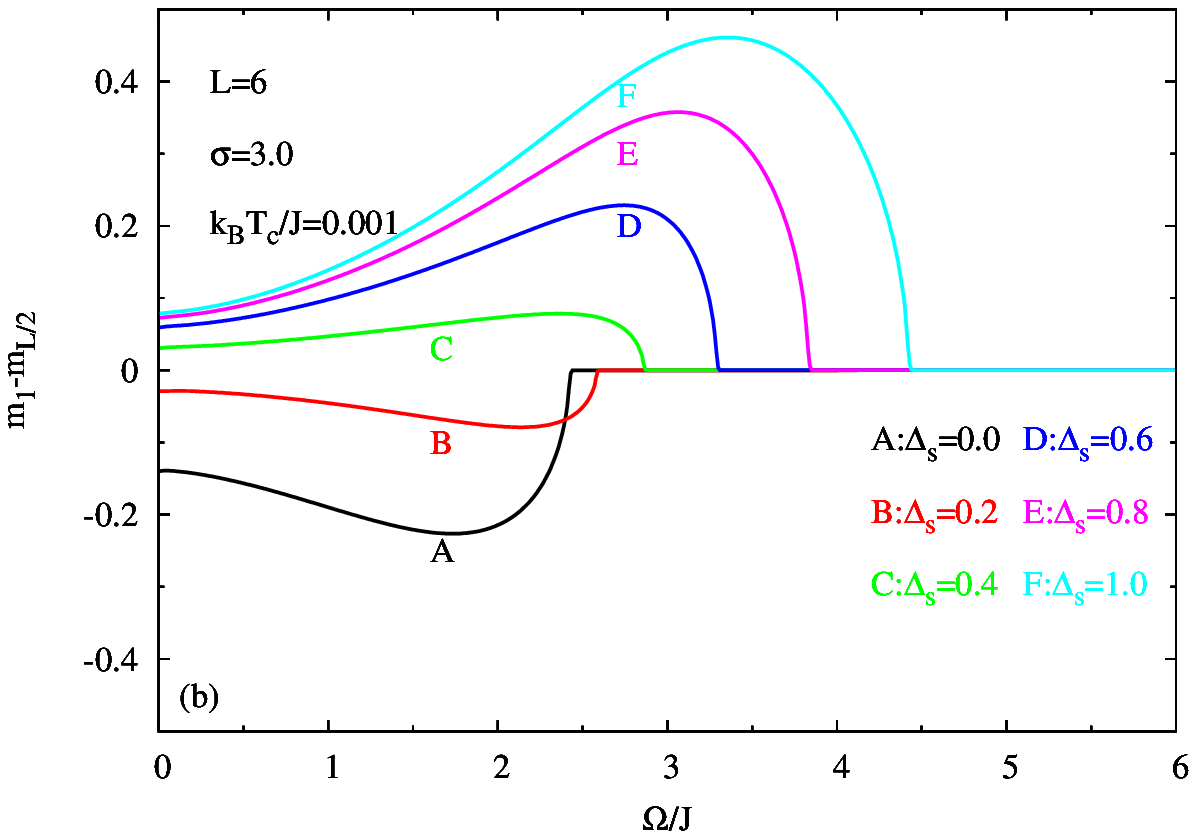, width=6cm}
\epsfig{file=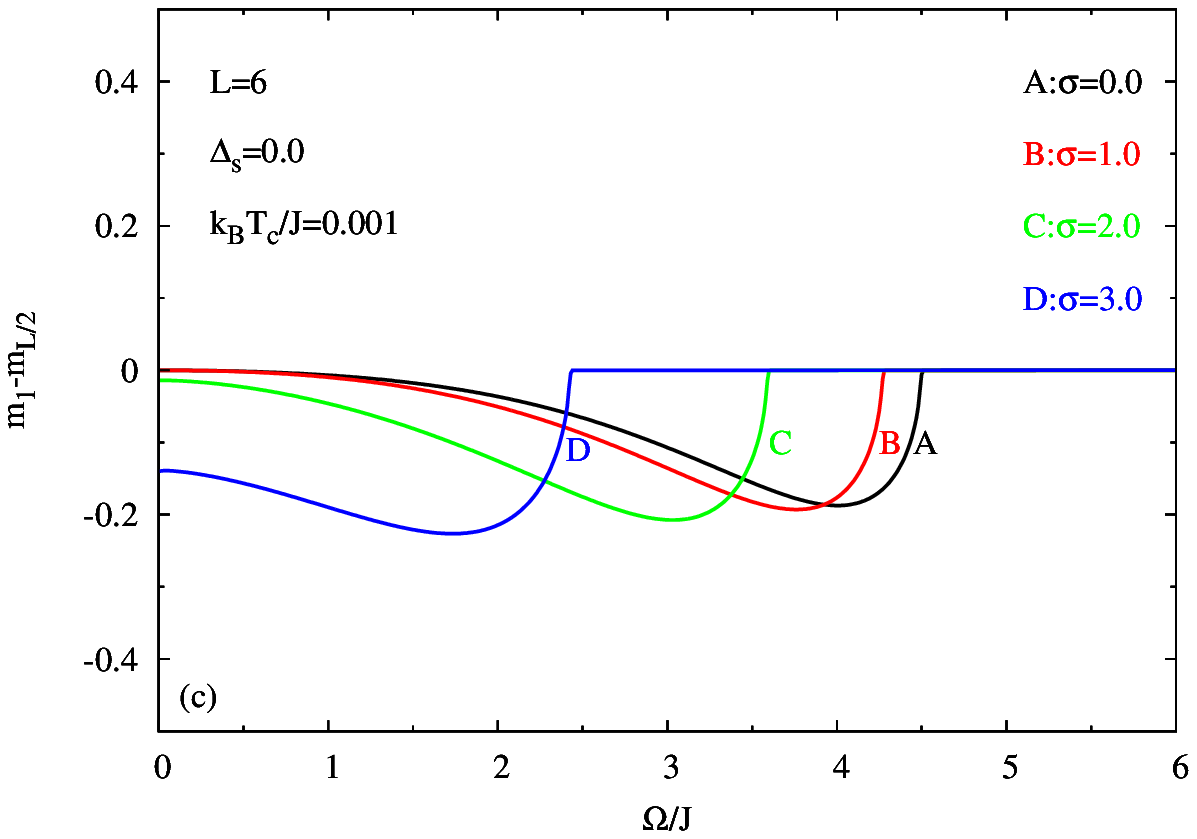, width=6cm}
\epsfig{file=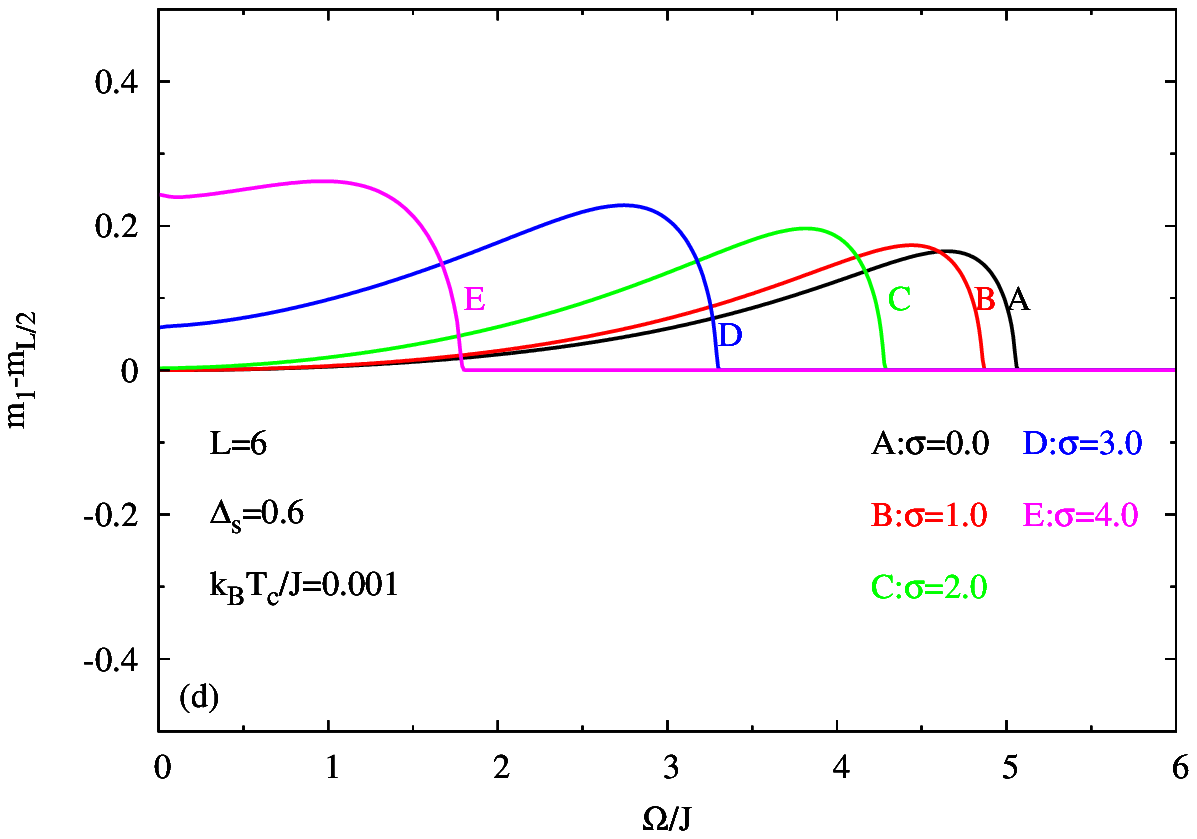, width=6cm}
\epsfig{file=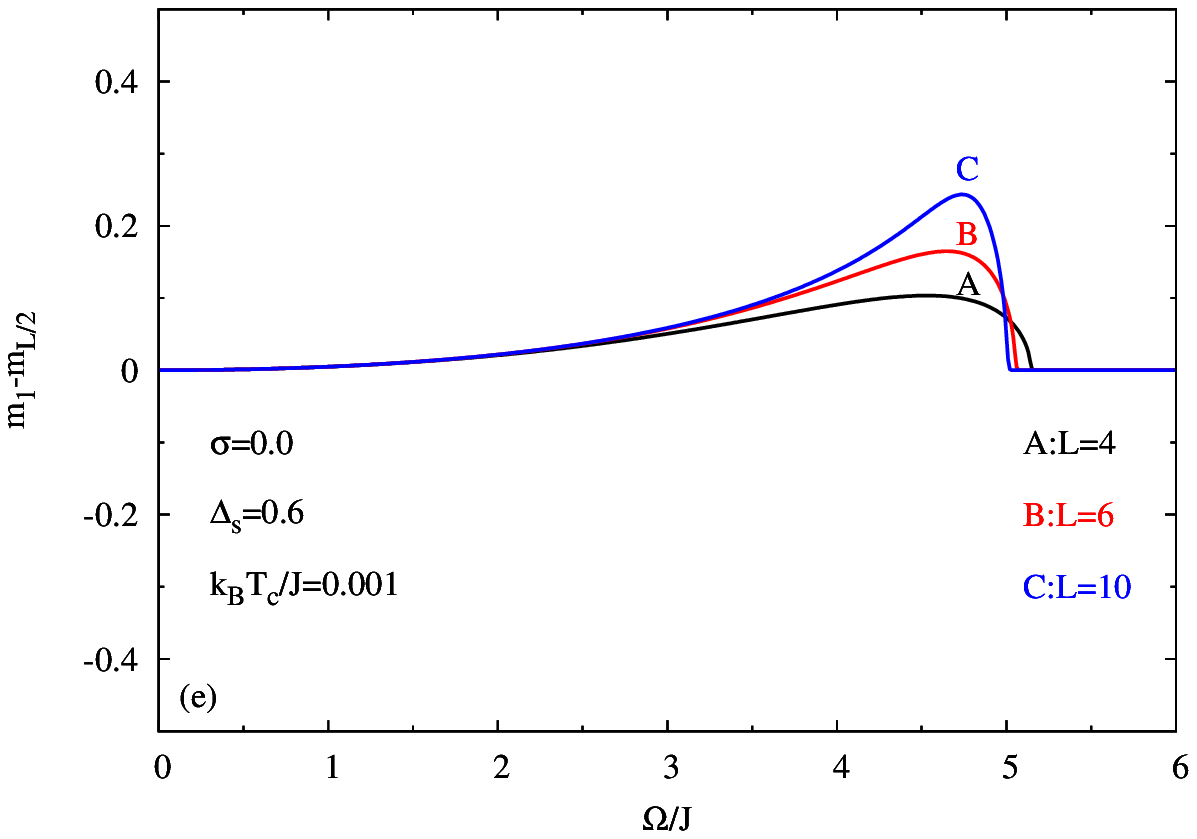, width=6cm}
\epsfig{file=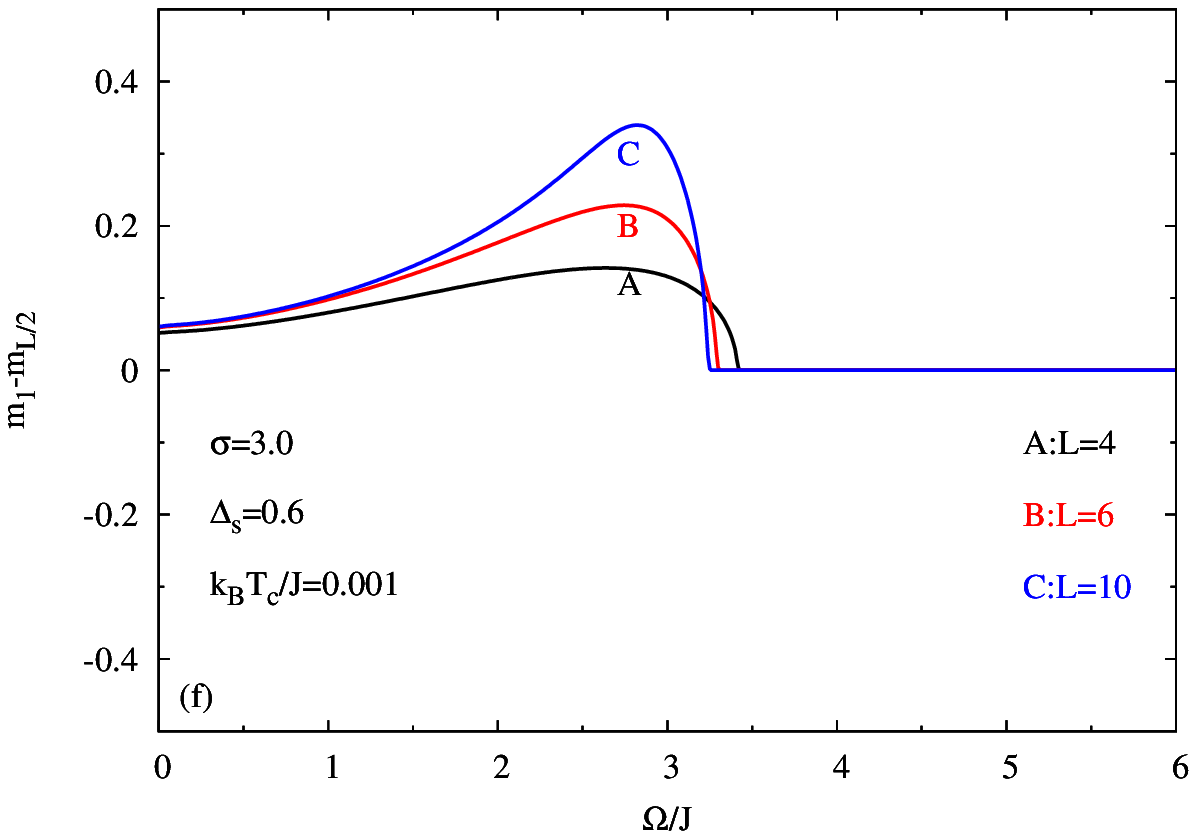, width=6cm}
\end{center}
\caption{Variation of the difference between the longitudinal magnetization of the surface  layer and the inner layer ($\Delta m=m_1-m_{L/2}$) calculated at the temperature $k_BT_c/J=0.001$  with the transverse field $\Omega/J$
for some selected values of $\Delta_s,L$ and $\sigma$ for the thin film described by the TIM  with single Gaussian random field distribution. The chosen value of $r_2$ is $r_2=1$.} \label{sek7}\end{figure}

\begin{figure}[h]\begin{center}
\epsfig{file=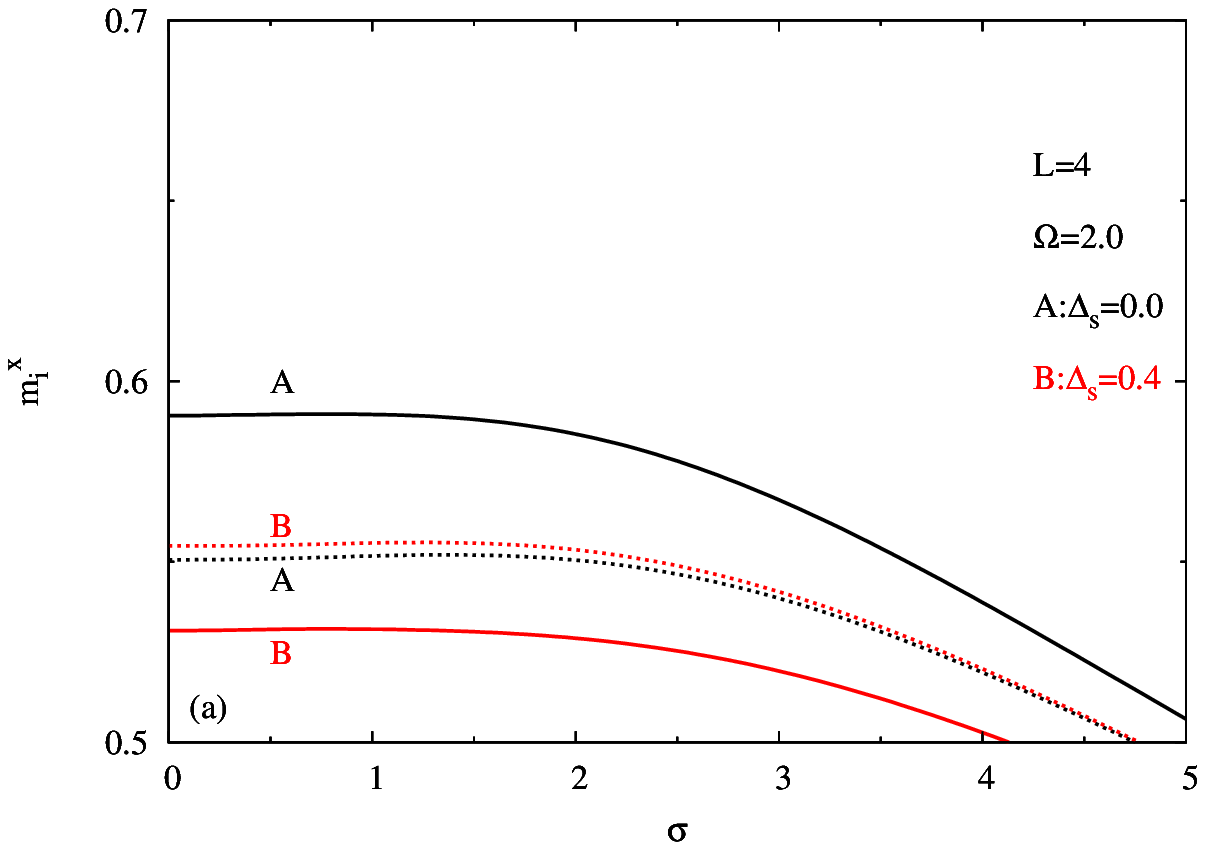, width=6cm}
\epsfig{file=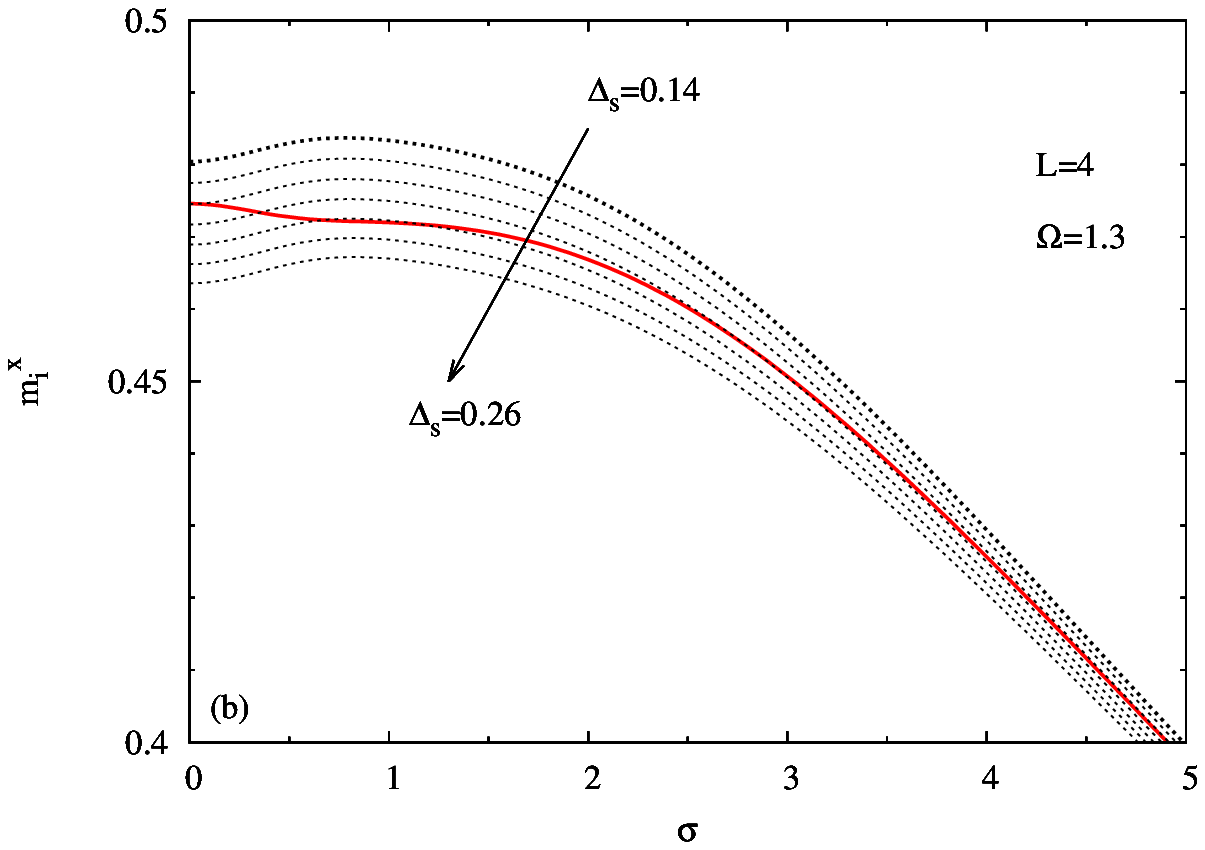, width=6cm}
\end{center}
\caption{(a): Variation of the surface layer transverse magnetization ($m_1^x$, solid line) and inner layer transverse magnetization ($m_{L/2}^x$, dotted line) with $\sigma$ evaluated at the temperature $k_BT_c/J=0.001$ for the values of $\Delta_s=0.0$ and $\Delta_s=0.4$. Fixed values are $r_2=1,L=4$ and $\Omega=2.0$. (b):  Variation of the surface layer transverse magnetization ($m_1^x$, dotted line) and inner layer transverse magnetization ($m_{L/2}^x$, solid line) with $\sigma$ evaluated at the temperature $k_BT_c/J=0.001$ for the values of from $\Delta_s=0.14$ to  $\Delta_s=0.26$. The increment of $\Delta_s$ from one curve to the neighbor curve is $0.02$. Fixed values are $r_2=1,L=4$ and $\Omega=1.3$.}
\label{sek8}\end{figure}

\begin{figure}[h]\begin{center}
\epsfig{file=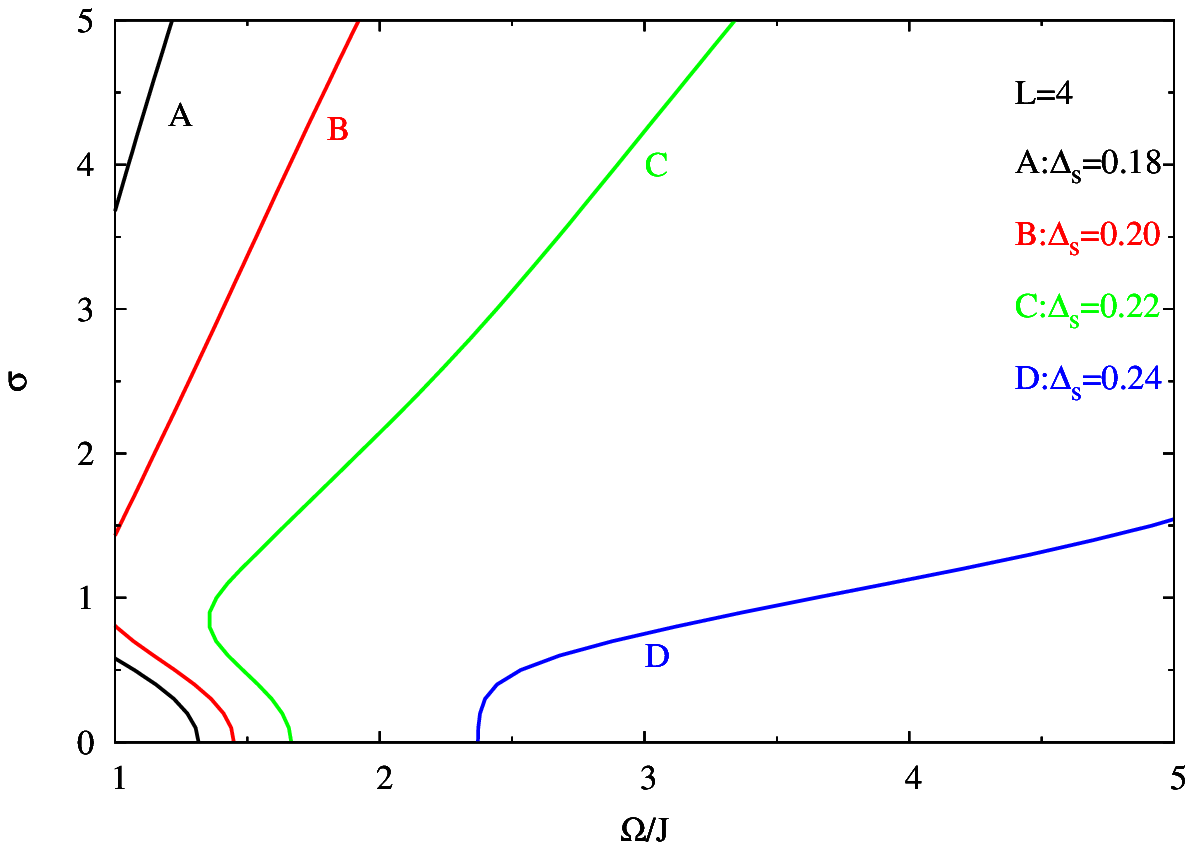, width=6cm}
\end{center}
\caption{
The borders of the positive valued $\Delta m^x$ and negative valued $\Delta m^x$ calculated at the temperature $k_BT_c/J=0.001$ for the film thickness $L=4$ and for selected values of $\Delta_s=0.18,0.20,0.22,0.24$ in the $(\sigma,\Omega/J)$ plane.
 $\Delta m^x$ is positive (i.e. $m_{1}^x>m_{L/2}^x$) on the region that covers the right side of the related $\Delta_s$ curve and negative (i.e. $m_{1}^x<m_{L/2}^x$) on the region that covers the left side of the related $\Delta_s$ curve.}
\label{sek9}\end{figure}

\section{Conclusion}\label{conclusion}

In this work, the effect of the zero centered Gaussian random magnetic field distribution on the phase
diagrams and ground state magnetizations of the  thin film described by the TIM  has been investigated.  As a formulation, the differential operator technique and DA within the  EFT has been used.

First, the evolution of the phase diagrams of the  thin film in the $(k_BT_c/J,\Delta_s)$, $(\Omega_c/J,\Delta_s)$ and $(k_BT_c/J,\Omega/J)$ planes with random Gaussian longitudinal magnetic field distribution width ($\sigma$) investigated in detail. Also the effect of the $\sigma$ on the special point which appears in the phase diagrams in the $(k_BT_c/J,\Delta_s)$ and  $(\Omega_c/J,\Delta_s)$ planes has been investigated. For the coordinates of the special point $(\Delta_s^{*},k_BT_c^{*}/J)$ and $(\Delta_s^*,\Omega_c^*/J)$ in the $(k_BT_c/J,\Delta_s)$, $(\Omega_c/J,\Delta_s)$  planes, respectively; we found that rising $\sigma$ makes no significant change in the $\Delta_s^{*}$ in both planes, but gives rise to a decline in the $k_BT_c^{*}/J$ and $\Omega_c^*/J$. After the value of $\sigma=4.22$, the special point disappears in both planes simultaneously. Although the special point can not appear for the values  $\sigma>4.22$, thin film can achieve ordered phase after some value of $\Delta_s$ as seen in Figs. \ref{sek1} (f) and \ref{sek2} (f). This region shows itself in the corresponding semi-infinite system with ordered surface and disordered bulk. At the same time, the effect of the rising $\sigma$  on the relation between the critical temperature/transverse field and film thickness has been discussed in detail.

Although there are different physical mechanisms underlying the observations, both rising $\Omega$ and $\sigma$ give rise to a decline in the critical temperature. In order to focus on the effect of $\sigma$ on the phase transition in the presence of transverse field, the variation of the ground state longitudinal/transverse magnetizations with $\sigma$ has been obtained. After making some conclusions on the effect of the rising $\sigma$ on the curves that represents the evolution of the magnetization with temperature,  we have focused on the effect of the $\sigma$ on the ground state magnetizations of the surface and inner layer. It is not surprising that, for small $\Delta_s$, the surface ground state transverse magnetization is greater than the inner one in the absence of the magnetic field. As explained in the Sec. \ref{results}, rising $\Delta_s$ reverses this situation, i.e. in terms of the defined parameter in the Sec. \ref{results}, $\Delta m^x=m_{1}^x-m_{L/2}^x<0$  which is calculated at the temperature $k_BT_c/J=0.001$. It has been shown that, within the $(\sigma,\Omega/J)$ plane,  for some values of the $\Delta_s$, we get $\Delta m^x>0$ while for the higher values we always have $\Delta m^x<0$ regardless of $\sigma$ and $\Omega$ value. However, in the region that covers the transition between the two cases, there is a complicated situation which comes from the competition between $\Omega$ and $\sigma$.

We hope that the results  obtained in this work may be beneficial form both theoretical and experimental point of view.

\end{document}